\documentclass[twocolumn,showpacs,preprintnumbers,amsmath,amssymb]{revtex4}

\usepackage{graphicx}
\usepackage{dcolumn}
\usepackage{bm}

\newcommand{\opr}[1]{\ensuremath{\mathbf{\mathsf{#1}}}}

\newcommand{\ket}[1]{\ensuremath{\left|#1\!\left.\right>\right.}}
\newcommand{\bra}[1]{\ensuremath{\left<#1\!\left.\right|\right.}}

\begin{document}
\date{\today}
\title{Characterizing Multiple Solutions to the Time - Energy Canonical Commutation Relation via Internal Symmetries}


\author{Roland Cristopher F. Caballar}
\email{rcaballar@nip.upd.edu.ph}
\author{Leonard R. Ocampo}
\author{Eric A. Galapon}
\email{eric.galapon@up.edu.ph}
\affiliation{Theoretical Physics Group, National Institute of Physics\\University of the Philippinies, Diliman, Quezon City 1101}
\begin{abstract}
Internal symmetries can be used to classify multiple solutions to the time energy canonical commutation relation (TE-CCR). The dynamical behavior of solutions to the TE-CCR posessing particular internal symmetries involving time reversal differ significantly from solutions to the TE-CCR without those particular symmetries, implying a connection between the internal symmetries of a quantum system, its internal unitary dynamics, and the TE-CCR.
\end{abstract}


\maketitle

\section{Introduction}
The status of time as a quantum mechanical observable has attracted, and continues to attract, much attention and controversy. Today, the status of time as a quantum observable continues to remain controversial due to a lack of a generalized approach in representing quantum time observables. As proof, numerous quantum time observables corresponding to a given physical system have been constructed using different procedures; examples of such quantum time observables are provided in references \cite{aharonov,allcock, kijowski, misra, srinivas, buttiker, sokolovski, hartle, yamada, bsm, busch1, schulman, halliwell, blanchard, grot, gian, pegg, muga, muga2, muga3, anas}. 

However, a generalized approach towards the representation of time as a quantum observable is detailed in references \cite{gal021, gal06}. In this approach, time is represented as an operator which satisfies the time-energy canonical commutation relation (TE-CCR) $\left(\opr{T}\opr{H}-\opr{H}\opr{T}\right)\ket{\phi}=\pm i\hbar\ket{\phi}$, where $\opr{H}$ is the Hamiltonian operator corresponding to the system in which the TE-CCR is formulated, and $\opr{T}$ is a time operator, or a solution to the TE-CCR. This approach implies that one can obtain a multitude of quantum time observables for a physical system simply by solving the TE-CCR. Applications of this approach towards the construction of quantum time observables for a particular physical system are detailed in references \cite{gal022,gal04,gal041,gal05,gal06a,gal08,gal09}. The time observables constructed in these references have attracted interest for both mathematical and physical reasons. In particular, the time operator constructed in reference \cite{gal022} has been shown by Arai \cite{arai} to be a generalized time operator, as well as relevant to the perturbation expansion of the system Hamiltonian. 

This generalized approach towards constructing quantum time observables leads to a multitude of quantum time observables for a given system, giving rise to a need for a mechanism to distinguish between these quantum time observables. Such a possible mechanism has been detailed in reference \cite{caballar}, where it was shown that one can distinguish between multiple solutions to the TE-CCR using their corresponding canonical domains, which are subspaces of $\mathcal{H}$ within which the TE-CCR is valid, as well as the system's internal unitary dynamics. Such a mechanism for physically distinguishing between multiple solutions to the TE-CCR is significant, since it shows that there is a possible relationship between time and the internal unitary dynamics of the system, allowing for a better understanding of the nature of time as a quantum observable. However, there is the question of whether it is possible to distinguish between multiple solutions to the TE-CCR of the same category, i. e. multiple solutions to the TE-CCR with identical canonical domains. This question is significant, since a negative answer to this question implies that, from a physical point of view, multiple solutions to the TE-CCR of the same category are identical to each other, so one can use any of those solutions to the TE-CCR to represent the quantum time observable corresponding to those solutions.

To be able to answer this question, we construct solutions to the TE-CCR of closed and dense category respectively, and show that the resulting solutions to the TE-CCR may be distinguished from each other using particular internal symmetries involving time reversal. We then use the methods first described in reference \cite{caballar} in order to determine whether one will be able to distinguish between multiple solutions to the TE-CCR of the same category via the system's internal unitary dynamics. The results of these investigations, which will be presented later in the paper, then imply that indeed, it is possible to distinguish between multiple solutions to the TE-CCR of the same category using not just the system's internal unitary dynamics, but also using certain internal symmetries which involve time reversal.

This paper represents our continuing efforts to understand the nature of time as a quantum observable. The rest of this paper is organized as follows. In section II, we give an overview of previous work done concerning multiple solutions to the TE-CCR of different categories. In sections III and IV, we show how multiple solutions to the TE-CCR of the same category can be mathematically and physically distinguishedfrom each other using particular internal symmetries involving time reversal and the system's internal unitary dynamics. We summarize our findings in section V.

\section{The Time - Energy Canonical Commutation Relation}
Based on discussions on the mathematical properties of canonical commutation relations in reference \cite{gal06}, the TE-CCR is valid only within a subspace $\mathcal{D}_{C}$ of the Hilbert space $\mathcal{H}$, where $\mathcal{D}_{C}$ is the canonical domain of the TE-CCR, which may either be dense or non-dense. If $\mathcal{D}_{C}$ is dense, then the time operator corresponding to $\mathcal{D}_{C}$ is said to be a solution to the TE-CCR of dense category. If, on the other hand, $\mathcal{D}_{C}$ is non-dense, then its closure is a proper subspace of $\mathcal{H}$, so $\mathcal{D}_{C}$ is said to be closed, and the time operator corresponding to $\mathcal{D}_{C}$ is said to be a solution to the TE-CCR of closed category.

A particular system in which a TE-CCR can be formulated and which was first described in references \cite{gal04,gal05} is a structureless particle of mass $\mu$ confined within a potential-free segment of the real line with endpoints at $-l$ and $l$, $l>0$. The Hilbert space $\mathcal{H}=\mathrm{L}^{2}[-l,l]$ is attached to this system. The Hamiltonian for this system is defined as $\opr{H}_{\gamma}=(2\mu)^{-1}\opr{p}_{\gamma}^{2}$, where $\opr{p}_{\gamma}=-i\hbar\partial_{q}$ is a self-adjoint momentum operator whose domain is comprised of all elements$\varphi(q)$ of $\mathcal{H}$ whose derivatives are square integrable and which satisfy the boundary conditions $\varphi(-l)=e^{-2i\gamma}\varphi(l)$. We note that the eigenfunctions of $\opr{H}_{\gamma}$ have the explicit form $\phi_{k}^{\gamma}(q)=(2l)^{-1/2}\exp\left[\frac{iq}{l}(\gamma+k\pi)\right]$ and their corresponding eigenvalues have the form $E_{k,\gamma}=(2\mu)^{-1}\left(\hbar l^{-1}\right)^{2}(\gamma+k\pi)^{2}$.

In reference \cite{gal06}, it has been shown that multiple solutions to the TE-CCR formulated in the system described in the previous paragraph exist, with nothing to mathematically forbid their existence. Two such solutions to this TE-CCR which were considered in reference \cite{gal06} are the characteristic time operator (CTO) $\opr{T}_{\gamma,2}$, which is of dense category and which was first presented in reference \cite{gal022}, and the confined quantum time of arrival operator (CTOA operator) $\opr{T}_{\gamma,1}$, which is of closed category and which was first presented in reference \cite{gal04}. These solutions to the TE-CCR were also shown in reference \cite{caballar} to be distinct from each other with respect to the system's internal unitary dynamics. As such, one can then distinguish between multiple solutions to the TE-CCR of different categories by means of either their canonical domain or the system's internal unitary dynamics. 

The use of the internal unitary dynamics to differentiate between the CTOA operator and the CTO has its origins in the use of the internal unitary dynamics to provide a physical interpretation for these time operators, a method which was first demonstrated in references \cite{gal04,gal05}. The standard way to interpret such operators would be via the quantum measurement postulate, which states that when a quantum observable is measured, the results of such a measurement of the observable will be one of the eigenvalues of the operator which represents the quantum observable in Hilbert space. Furthermore, such a measurement may be taken at any instant of time. In the case of a quantum time operator then, its eigenvalues will have units of time, and there can be countably many results obtained in taking a quantum measurement of a quantum time observable. A problem then occurs when we consider that the quantum measurement can be taken at any instant of time. The problem occurs because, as stated earlier, any one of the countably many eigenvalues of the quantum time operator corresponding to the quantum time observable being measured can emerge as the result of measuring such an observable. So it is possible then that at an instant of time equal to 15 seconds, the result of a quantum measurement of a quantum time observable will be one of the eigenvalues of the operator representation of that quantum observable whose magnitude is 3 seconds, in which case the quantum measurement postulate is inadequate in resolving this apparent paradox. Thus there is a need to devise a mechanism to physically interpret these time operators without having to resort to the use of the quantum measurement postulate, a mechanism provided by the system's internal unitary dynamics.

However, there now arises the question of how one can distinguish between multiple solutions to the TE-CCR of the same category, i. e. multiple solutions to the TE-CCR with identical canonical domains. To be able to answer this question, there is a need to determine a mathematical property of a given solution to the TE-CCR which is independent of the solution's canonical domain; this mathematical property can then be utilized to distinguish between multiple solutions to the TE-CCR of the same category. There is also a need to determine whether multiple solutions to the TE-CCR of the same category have identical dynamical behaviors. If a mathematical property is found that can be utilized to distinguish between multiple soltuions to the TE-CCR of the same category, and if multiple solutions to the TE-CCR of the same category do not have identical dynamical behaviors, then one can distinguish between multiple solutions to the TE-CCR not just using a certain mathematical property, but also by means of the system's internal unitary dynamics. We will attempt to answer this question in the next two sections of the paper.

\section{Existence of Multiple Solutions to the TE-CCR of Closed Category}
Let us first consider multiple solutions to the TE-CCR described in the previous section, all of which which have non-dense, or closed, canonical domains. One possible solution to the TE-CCR of closed category has been described in references \cite{gal04,gal05}. This solution to the TE-CCR is known as the confined quantum time of arrival (CTOA) operator, constructed via Weyl quantization of the classical arrival time $t=-\mu qp^{-1}$ for this system. It is this solution to the TE-CCR that we use to generate other solutions to the TE-CCR of closed category.

To see how the CTOA operator can be used to generate other solutions to the TE-CCR of closed category, let us consider the classical arrival time for a given particle defined in an arbitrary physical system. The most general quantization of the classical arrival time for a given particle will result in the following operator:
\begin{equation}
\opr{T}=-\mu\frac{\opr{q}\opr{p}^{-1}+\opr{p}^{-1}\opr{q}+is(\opr{q}\opr{p}^{-1}-\opr{p}^{-1}\opr{q})}{2}
\label{ato1}
\end{equation}
where $\opr{q}$ and $\opr{p}^{-1}$ are the position and inverse momentum operators for the system, and $s$ is a real number denoting the method of quantization used. Now for the physical system described in the previous section, equation \ref{ato1} can be expressed in position representation as follows:
\begin{equation}
\left(\opr{T}_{\gamma,s,1}\varphi\right)(q)=\int_{-l}^{l}\left\langle q\right|\opr{T}_{\gamma,s,1}\left|q'\right\rangle T_{\gamma,s,1}(q,q')\varphi(q')dq'
\label{ato2}
\end{equation}
for $\gamma\neq 0$, where $\left\langle q\right|\opr{T}_{\gamma,s,1}\left|q'\right\rangle$ has the form
\begin{eqnarray}
&&\left\langle q\right|\opr{T}_{\gamma,s,1}\left|q'\right\rangle=-\frac{\mu}{4\hbar\sin\gamma}\times\nonumber\\
&&\left(e^{i\gamma}\mathrm{H}(q-q')+e^{-i\gamma}\mathrm{H}(q'-q)\right)((q+q')+is(q-q'))\nonumber\\
\label{atokern}
\end{eqnarray}
and
\begin{eqnarray}
&&\left(\opr{T}_{0,s,1}\varphi\right)(q)=\int_{-l}^{l}\left\langle q\right|\opr{T}_{0,s,1}\left|q'\right\rangle\varphi(q')dq'\nonumber\\
\label{ato2gammazero}
\end{eqnarray}
for $\gamma=0$, and where $\left\langle q\right|\opr{T}_{0,s,1}\left|q'\right\rangle$ has the form
\begin{eqnarray}
&&\left\langle q\right|\opr{T}_{0,s,1}\left|q'\right\rangle=-\frac{\mu i}{4\hbar}((q+q')+is(q-q'))\mathrm{sgn}(q-q')\nonumber\\
&&+\frac{\mu i}{4\hbar l}\left((q^{2}-q'^{2})-is(q-q')^{2}\right)
\label{atokerngammazero}
\end{eqnarray}
We note that equation \ref{atokerngammazero} was first stated in reference \cite{bsthesis}. The first terms in equations \ref{atokern} and \ref{atokerngammazero} are just the kernels for the CTOA operators $\opr{T}_{\gamma,1}$ and $\opr{T}_{0,1}$ respectively, and correspond to the term $2^{-1}\mu\left(\opr{q}\opr{p}^{-1}+\opr{p}^{-1}\opr{q}\right)$ in equation \ref{ato1}. The second term in equations \ref{atokern} and \ref{atokerngammazero}, on the other hand, correspond to the term $2^{-1}is\mu\left(\opr{q}\opr{p}^{-1}-\opr{p}^{-1}\opr{q}\right)$ in equation \ref{ato1} which in turn commutes with the Hamiltonian $\opr{H}_\gamma$. It can be shown that equations \ref{atokern} and \ref{atokerngammazero} are square integrable, which implies that the operators $\opr{T}_{\gamma,s,1}$ and $\opr{T}_{0,s,1}$ are self-adjoint and compact. As such, the operators $\opr{T}_{\gamma,s,1}$ and $\opr{T}_{\gamma,s,1}$ are constructed by adding to the CTOA operators $\opr{T}_{\gamma,1}$ and $\opr{T}_{0,1}$ a term which commutes with the Hamiltonian, but which at the same time ensures the self-adjointness and compactness of the resulting operators.

Since the second term of the operators $\opr{T}_{\gamma,s,1}$ and $\opr{T}_{0,s,1}$ commute with the Hamiltonian, the canonical domains of the operators $\opr{T}_{\gamma,s,1}$ and $\opr{T}_{\gamma,1}$ are identical, and so are the canonical domains of the operators $\opr{T}_{0,s,1}$ and $\opr{T}_{0,1}$. And since $s$ is a real number, there are an infinite number of solutions to the TE-CCR $\opr{T}_{\gamma,s,1}$ and $\opr{T}_{0,s,1}$ that can exist, each of which can be distinguished from each other by means of their corresponding values of $s$. However, is it still possible to distinguish, either mathematically or physically, between multiple solutions to the TE-CCR $\opr{T}_{\gamma,s,1}$ without having to determine the exact value of the parameter $s$ corresponding to them? We will answer this question in the next subsection, and in so doing we will be able to clarify the role of the parameter $s$ corresponding to the solution to the TE-CCR $\opr{T}_{\gamma,s,1}$ of closed category.

\subsection{Use of Internal Symmetries to Distinguish Between Multiple Solutions to the TE-CCR}
The role of the parameter $s$ in $\opr{T}_{\gamma,s,1}$ can be determined by examining the internal symmetries satisfied by $\opr{T}_{\gamma,s,1}$. It was shown in references \cite{gal04,gal05} that the CTOA operator $\opr{T}_{\gamma,1}$ satisfies the following internal symmetries:
\begin{eqnarray}
&&\opr{\Pi}^{-1}\opr{\Theta}^{-1}\opr{T}_{\gamma,1}\opr{\Theta}\opr{\Pi}=-\opr{T}_{\gamma}\nonumber\\
&&\opr{\Theta}^{-1}\opr{T}_{\gamma,1}\opr{\Theta}=-\opr{T}_{-\gamma}\nonumber\\
&&\opr{\Pi}^{-1}\opr{T}_{\gamma,1}\opr{\Pi}=\opr{T}_{-\gamma}\nonumber\\
\label{intsymm}
\end{eqnarray}
where $\opr{\Pi}$ is the parity reversal operator, $\left(\opr{\Pi}\varphi\right)(q)\varphi(-q)=$ and $\opr{\Theta}$ is the time reversal operator, $\left(\opr{\Theta}\varphi\right)(q)=\varphi*(q)$. The physical significance of these internal symmetries were discussed in reference \cite{gal05}. It can be shown that none of these symmetries are satisfied by $\opr{T}_{\gamma,s,1}$if $s\neq 0$, due to the presence of the second term in equation \ref{atokern}. As such, the parameter $s$ corresponding to $\opr{T}_{\gamma,s,1}$ serves to break three particular internal symmetries for $\opr{T}_{\gamma,s,1}$, with the explicit form of those symmetries given by equation \ref{intsymm}, and we can then mathematically distinguish between $\opr{T}_{\gamma,s,1}$ and $\opr{T}_{\gamma,1}$ in terms of the internal symmetries specified in equation \ref{intsymm}.

It was also shown in references \cite{gal04,gal05} that the CTOA operator $\opr{T}_{0,1}$ satisfies the following internal symmetries:
\begin{eqnarray}
&&\opr{\Theta}^{-1}\opr{T}_{0,1}\opr{\Theta}=-\opr{T}_{0}\nonumber\\
&&\opr{\Pi}^{-1}\opr{T}_{0,1}\opr{\Pi}=\opr{T}_{0}\nonumber\\
\label{intsymm2}
\end{eqnarray}
The physical significance of these internal symmetries are discussed in reference \cite{gal05}. None of these symmetries are satisfied by $\opr{T}_{0,s,1}$, again due to the presence of the second term in equation \ref{atokerngammazero}. So the role of $s$ as an internal symmetry breaking parameter is further underscored, and we then see that the CTOA operator $\opr{T}_{0,1}$ differs from the solution to the TE-CCR $\opr{T}_{0,s,1}$ respectively with respect to the internal symmetry given in equation \ref{intsymm2}.

We pay particular attention to the second internal symmetry specified in equation \ref{intsymm} and the first internal symmetry specified in equation \ref{intsymm2}. Those internal symmetries, which involve the use of the time reversal operator alone, correspond to the following symmetry obeyed by the classical arrival time:
\begin{equation}
-t(q,p)=t(q,-p)
\label{symmclass} 
\end{equation}
We note that equation \ref{symmclass} implies that a reversal of the classical arrival time can be carried out by a reversal of momentum. However, the second internal symmetry specified in equation \ref{intsymm} implies that due to the presence of the phase factor $\gamma$, in order for a full time reversal to occur, the phase must be reversed as well, since only by reversing the phase can the momentum be reversed. Because of this additional phase reversal operation, the second internal symmetry in equation \ref{intsymm} cannot be considered as a full time reversal symmetry despite it being the quantum analogue of equation \ref{symmclass}. Instead, we call this internal symmetry $\tau$-symmetry. However, the first internal symmetry in equation \ref{intsymm2} does not involve a reversal of the phase factor $\gamma$ in order for full time reversal to take place, and merely requires that the momentum be reversed; this internal symmetry is then said to be a proper time reversal operation. As such, we then say that the CTOA operator $\opr{T}_{\gamma,1}$ is $\tau$-symmetric, whereas the solution to the TE-CCR $\opr{T}_{\gamma,s,1}$ is non $\tau$-symmetric. Similarly, we say that the CTOA operator $\opr{T}_{0,1}$ is time reversal symmetric, whereas the solution to the TE-CCR $\opr{T}_{\gamma,s,1}$ is non time reversal symmetric. We note that time reversal symmetry is a special case of $\tau$-symmetry, which can only be seen when $\gamma=0$.

\subsection{Physical Implications of $\tau$-symmetry}
We have shown in the previous section that it is possible to distinguish between the solutions to the TE-CCR $\opr{T}_{\gamma,1}$ and $\opr{T}_{\gamma,s,1}$ as well as the solutions to the TE-CCR $\opr{T}_{0,1}$ and $\opr{T}_{0,s,1}$ via the internal symmetries given by equations \ref{intsymm} and \ref{intsymm2} respectively. An effect of the lack of $\tau$-symmetry in $\opr{T}_{\gamma,s,1}$ is that unlike the CTOA operator, all positive eigenvalues of $\opr{T}_{\gamma,s,1}$ will not have negative eigenvalue counterparts of equal magnitude; that is, if $\tau_{n}$ is a positive eigenvalue of $\opr{T}_{\gamma,s,1}$, then there will exist no negative eigenvalue $-\tau_{n'}$ of $\opr{T}_{\gamma,s,1}$ such that $\left|-\tau_{n'}\right|=\left|\tau_{n}\right|$. 

However, this is not just the only effect of the lack of $\tau$-symmetry in $\opr{T}_{\gamma,s,1}$. To be able to determine other effects of the lack of $\tau$-symmetry in $\opr{T}_{\gamma,s,1}$, we will compare the time evolution of the eigenfunctions of $\opr{T}_{\gamma,1}$ with the eigenfunctions of $\opr{T}_{\gamma,s,1}$, and the time evolution of the eigenfunctions of $\opr{T}_{0,s,1}$ with the eigenfunctions of $\opr{T}_{0,s,1}$.
\subsubsection{Computation of the Eigenfunctions and the Eigenvalues of $\opr{T}_{\gamma,s,1}$}
Before we can actually do so, however, let us first determine the explicit form of the eigenfunctions of $\opr{T}_{\gamma,s,1}$. First note that the eigenvalue equation for $\opr{T}_{\gamma,s,1}$ can be written in the following form:
\begin{eqnarray}
&&\varphi_{n,\gamma,\nu}^{s,1}(q)=-\frac{\mu}{4\tau_{n}\hbar\sin\gamma}\int_{-l}^{l}((q+q')+is(q-q'))\times\nonumber\\
&&\left(e^{i\gamma}\mathrm{H}(q-q')+e^{-i\gamma}\mathrm{H}(q'-q)\right)\varphi_{n,\gamma,\nu}^{s,1}(q')dq'\nonumber\\
\label{eigeneqato}
\end{eqnarray}
where $\varphi_{n,\gamma,\nu}^{s,1}$ are the eigenfunctions of $\opr{T}_{\gamma,s,1}$ and $\tau_n$ the corresponding eigenvalues. We solve for the eigenfunctions and eigenvalues of $\opr{T}_{\gamma,s,1}$ in the same manner that we solved for the eigenfunctions and eigenvalues of the CTOA operator in reference \cite{gal04,gal05}. In doing so, we find that the explicit form of the eigenvalues of $\opr{T}_{\gamma,s,1}$ will have the explicit form
\begin{eqnarray}
&&\varphi_{n,\gamma,\nu}^{s,1}(q)=A_{0}\;_{1}\mathrm{F}_{1}\left(\frac{3+is}{4};\frac{1}{2};-ir_{n}\frac{q^2}{l^2}\right)+\nonumber\\
&&\alpha_{1}q\;_{1}\mathrm{F}_{1}\left(\frac{5+is}{4};\frac{3}{2};-ir_{n}^{\nu}\frac{q^2}{l^2}\right)\nonumber\\
\label{nonperatoeigenfin}
\end{eqnarray}
where $_{1}\mathrm{F}_{1}(a;b;x)$ are hypergeometric functions, $A_0$ is the normalization constant, $r_{n}^{\nu}=\pm\frac{\mu l^2}{2\tau_{n}\hbar}$, with $\tau_n$ the $n$th eigenvalue of $\opr{T}_{\gamma,s,1}$, and
\begin{eqnarray}
\alpha_{1}=\frac{3A_{0}\;_{1}\mathrm{F}_{1}\left(\frac{3+is}{4};\frac{3}{2};-ir_{n}^{\nu}\right)(i-s)r_{n}^{\nu}\;}{l\tan\gamma\left(
\begin{array}{c}
r_{n}^{\nu}(1-is)\;_{1}\mathrm{F}_{1}\left(\frac{5+is}{4};\frac{5}{2};-ir_{n}^{\nu}\right)-\\
3i\;_{1}\mathrm{F}_{1}\left(\frac{5+is}{4};\frac{3}{2};-ir_{n}^{\nu}\right)
\end{array}
\right)}
\end{eqnarray}
On the other hand, the eigenvalues $\tau_{n}$ of $\opr{T}_{\gamma,s,1}$ are the roots of the characteristic equation $T_{11}T_{22}-T_{21}T_{12}=0$, where 
\begin{eqnarray}
&&T_{11}=_{1}\mathrm{F}_{1}\left(\frac{3+is}{4};\frac{1}{2};-ir_{n}^{\nu}\right)-\nonumber\\
&&\frac{r_{n}^{\nu}e^{-i\gamma}(1+is)}{\sin\gamma}\;_{1}\mathrm{F}_{1}\left(\frac{3+is}{4};\frac{3}{2};-ir_{n}^{\nu}\right)\nonumber\\
&&T_{12}=-l\;_{1}\mathrm{F}_{1}\left(\frac{5+is}{4};\frac{3}{2};-ir_{n}^{\nu}\right)+\nonumber\\
&&\frac{r_{n}^{\nu} e^{-i\gamma}l(1-is)}{3\sin\gamma}\;_{1}\mathrm{F}_{1}\left(\frac{5+is}{4};\frac{5}{2};-ir_{n}^{\nu}\right)\nonumber\\
&&T_{21}=_{1}\mathrm{F}_{1}\left(\frac{3+is}{4};\frac{1}{2};-ir_{n}^{\nu}\right)+\nonumber\\
&&\frac{r_{n}^{\nu}e^{i\gamma}(1+is)}{\sin\gamma}\;_{1}\mathrm{F}_{1}\left(\frac{3+is}{4};\frac{3}{2};-ir_{n}^{\nu}\right)\nonumber\\
&&T_{22}=l\;_{1}\mathrm{F}_{1}\left(\frac{5+is}{4};\frac{3}{2};-ir_{n}^{\nu}\right)+\nonumber\\
&&\frac{r_{n}^{\nu} e^{i\gamma}l(1-is)}{3\sin\gamma}\;_{1}\mathrm{F}_{1}\left(\frac{5+is}{4};\frac{5}{2};-ir_{n}^{\nu}\right)\nonumber\\
\label{matel}
\end{eqnarray}
\subsubsection{Special Case: $\opr{T}_{\pi/2,s,1}$}
We consider the operator $\opr{T}_{\pi/2,s,1}$ separately from $\opr{T}_{\gamma,s,1}$, where $\gamma\neq\pi/2$ as a special case, since when $\gamma$ assumes this value, the boundary conditions governing the elements of the domain of the Hamiltonian of the system assume the form $\varphi(-l)=-\varphi(l)$, which are antiperiodic. For this case, the eigenfunctions of $\opr{T}_{\pi/2,s,1}$ as given in equation \ref{nonperatoeigenfin}, bifurcate into odd and even parity components. Explicitly, these eigenfunctions will have the form
\begin{equation}
\varphi_{n,\pi/2,\nu}^{s,1,e}(q)=A_{0}\;_{1}\mathrm{F}_{1}\left(\frac{3+is}{4};\frac{1}{2};-ir_{n}^{\nu}\frac{q^2}{l^2}\right)
\label{eigenatoantipereven}
\end{equation}
\begin{equation}
\varphi_{n,\pi/2,\nu}^{s,1,o}(q)=qA_{1}\;_{1}\mathrm{F}_{1}\left(\frac{5+is}{4};\frac{3}{2};-ir_{n}^{\nu}\frac{q^2}{l^2}\right)
\label{eigenatoantiperodd}
\end{equation}
where $A_{0}$ and $A_{1}$ are normalization constants and the notations $e$ and $o$ in the superscript denoting whether the eigenfunction has even or odd parity, respectively. On the other hand, the characteristic equations corresponding to $\varphi_{n,\pi/2,\nu}^{s,e,1}(q)$ and $\varphi_{n,\pi/2,\nu}^{s,o,1}(q)$ respectively will have the following explicit form:
\begin{eqnarray}
&&-\frac{ir_{n}(1+is)}{l\sin\gamma}\;_{1}\mathrm{F}_{1}\left(\frac{3+is}{4};\frac{3}{2};-ir_{n}\right)+\nonumber\\
&&_{1}\mathrm{F}_{1}\left(\frac{3+is}{4};\frac{1}{2};-ir_{n}\right)=0
\label{chareqatoantipereven}
\end{eqnarray}
\begin{eqnarray}
&&\frac{ ilr_{n}(1-is)}{3\sin\gamma}\;_{1}\mathrm{F}_{1}\left(\frac{5+is}{4};\frac{5}{2};-ir_{n}\right)+\nonumber\\
&&l\;_{1}\mathrm{F}_{1}\left(\frac{5+is}{4};\frac{3}{2};-ir_{n}\right)=0
\label{chareqatoantiperodd}
\end{eqnarray}
The roots of these equations will then give us the eigenvalues corresponding to the eigenfunctions of $\opr{T}_{\pi/2,s,1}$.
\subsubsection{Computation of the Eigenfunctions and the Eigenvalues of $\opr{T}_{0,s,1}$}
For the solution to the TE-CCR $\opr{T}_{0,s,1}$, we use the same techniques used to solve for the eigenfunctions and eigenvalues of the CTOA operator $\opr{T}_{0,s,1}$. In doing so, we find that $\opr{T}_{0,s,1}$ will have two sets of linearly independent eigenfunctions. The explicit form of the first set, which is of odd parity and which we designate as $\varphi_{n,0,\nu}^{s,1,o}$, is identical to the odd parity eigenfunctions of $\opr{T}_{\pi/2,s,1}$, given by equation \ref{eigenatoantiperodd}. Furthermore, the characteristic equation which gives the eigenvalues corresponding to the odd parity eigenfunctions $\varphi_{n,0,\nu}^{s,1,o}(q)$ is identical to equation \ref{chareqatoantiperodd}. On the other hand, it can be shown that the second set of eigenfunctions of $\opr{T}_{0,s,1}$, which we designate as $\varphi_{n,0,\nu}^{s,1,e}(q)$, will have the following explicit form:
\begin{eqnarray}
&&\varphi_{n,0,\nu}^{s,1,e}(q)=A_{0}\mathrm{F}_{1}\left(\frac{3+is}{4};\frac{3}{2};-ir_{n}^{\nu}\frac{q^2}{l^2}\right)+\nonumber\\
&&\frac{2A_{0}(1-is)}{1+3is}\;_{1}\mathrm{F}_{1}\left(\frac{3+is}{4};\frac{3}{2};-ir_{n}^{\nu}\frac{q^2}{l^2}\right)\nonumber\\
\label{eigenzeroevenmod}
\end{eqnarray}
with $r_{n}^{\nu}=\pm\frac{\mu l^2}{2\tau_{n}\hbar}$ the roots of the characteristic equation
\begin{eqnarray}
&&\frac{\mu il^{2}(1+is)}{6\tau_{n}\hbar}\;_{P}\mathrm{F}_{Q}\left(\frac{3}{2},\frac{3+is}{4};\frac{1}{2},\frac{5}{2};-ir_{n}^{\nu}\right)\nonumber\\
&&+\frac{2(1-is)-2r_{n}^{\nu}s(3+is)+\frac{2ir_{n}^{\nu}}{3}\left(1+s^{2}\right)}{1+3is}\times\nonumber\\
&&\;_{1}\mathrm{F}_{1}\left(\frac{3+is}{4};\frac{3}{2};-ir_{n}^{\nu}\right)\nonumber\\
&&+_{1}\mathrm{F}_{1}\left(\frac{3+is}{4};\frac{1}{2};-ir_{n}^{\nu}\right)=0\nonumber\\
\label{chareqatopereven}
\end{eqnarray}
\subsubsection{Time Evolution of the Eigenfunctions of $\opr{T}_{\gamma,s,1}$}
Having computed for the eigenfunctions of $\opr{T}_{\gamma,s,1}$, we now proceed to evolve these eigenfunctions over time, in the same manner that the CTOA operator eigenfunctions were evolved. We present in this portion of the paper the results for the time evolution of the odd parity eigenfunctions of $\opr{T}_{\pi/2,s,1}$ as well as the time evolution of the even parity eigenfunctions of $\opr{T}_{0,s,1}$ only; similar results can be obtained when one evolves the other eigenfunctions of $\opr{T}_{\gamma,s,1}$. 

\begin{figure}
\includegraphics[width=0.23\textwidth,height=0.23\textwidth]{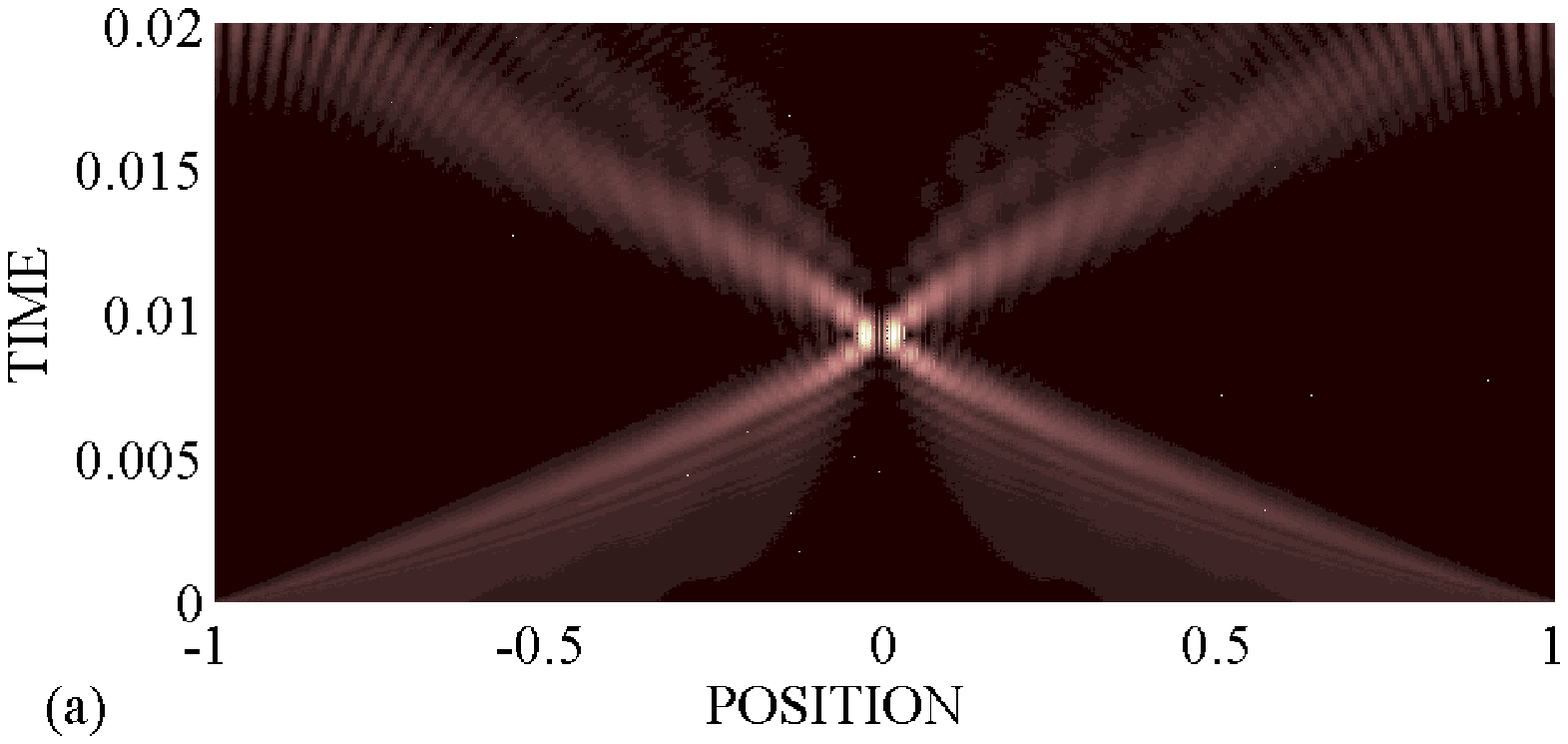}
\includegraphics[width=0.23\textwidth,height=0.23\textwidth]{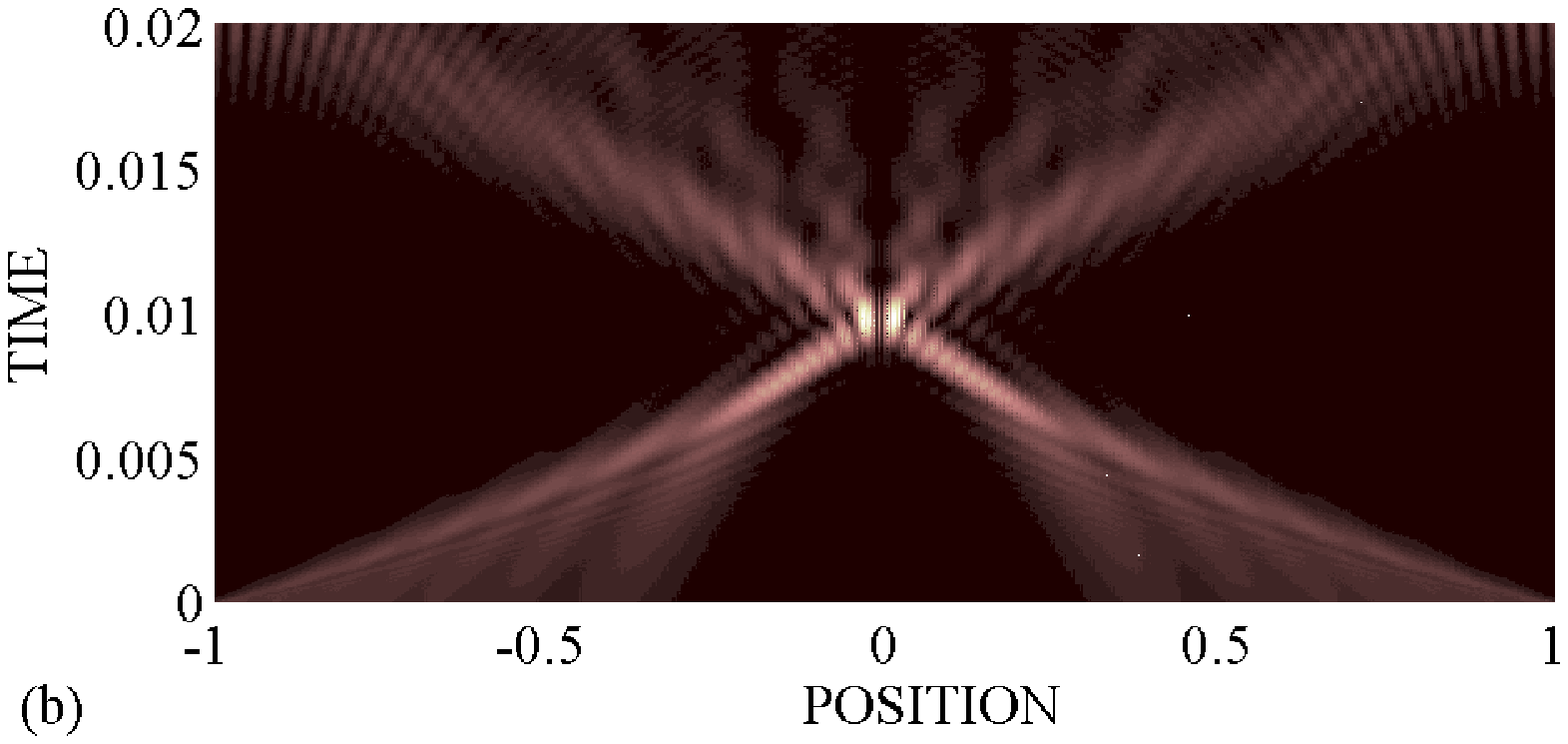}
\includegraphics[width=0.23\textwidth,height=0.23\textwidth]{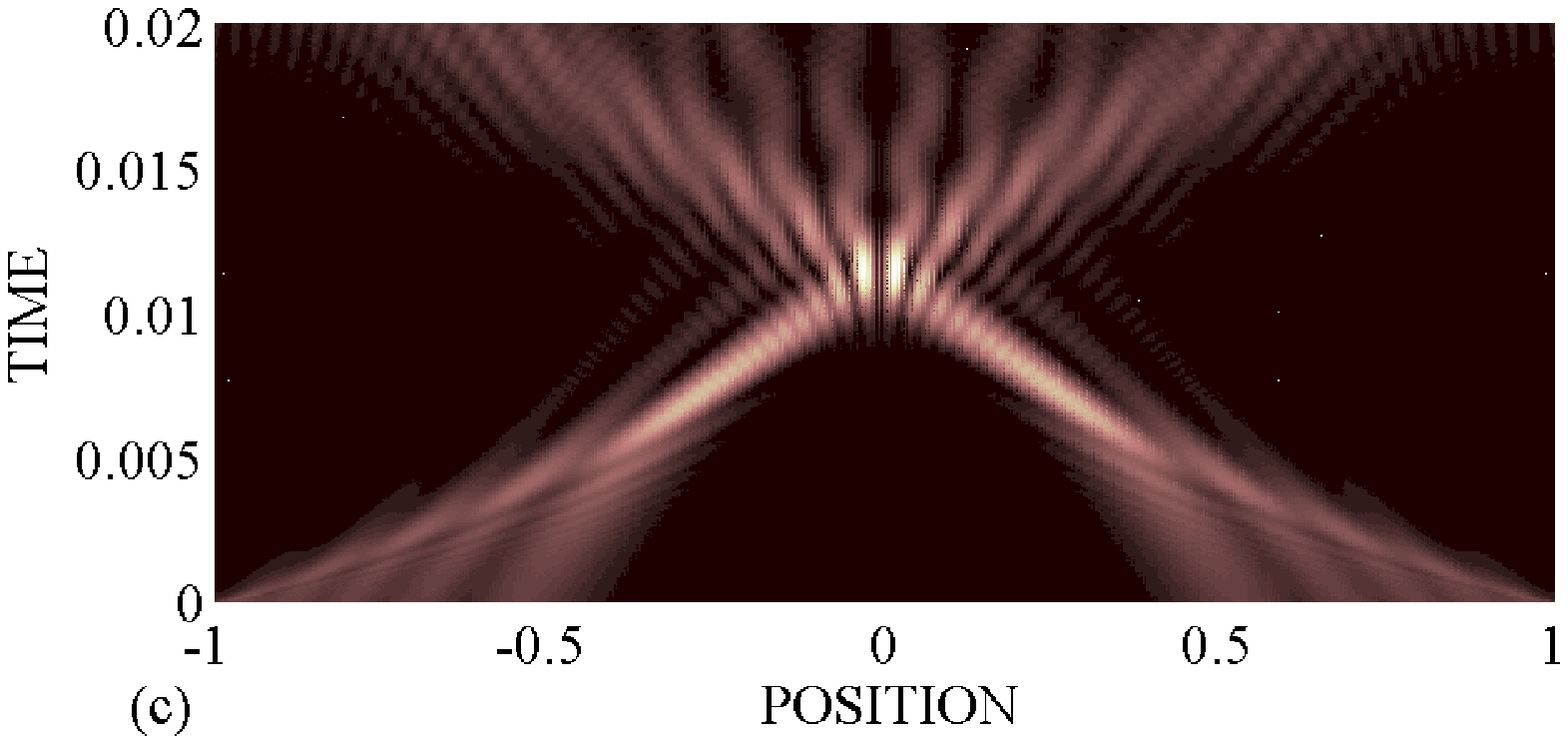}
\includegraphics[width=0.23\textwidth,height=0.23\textwidth]{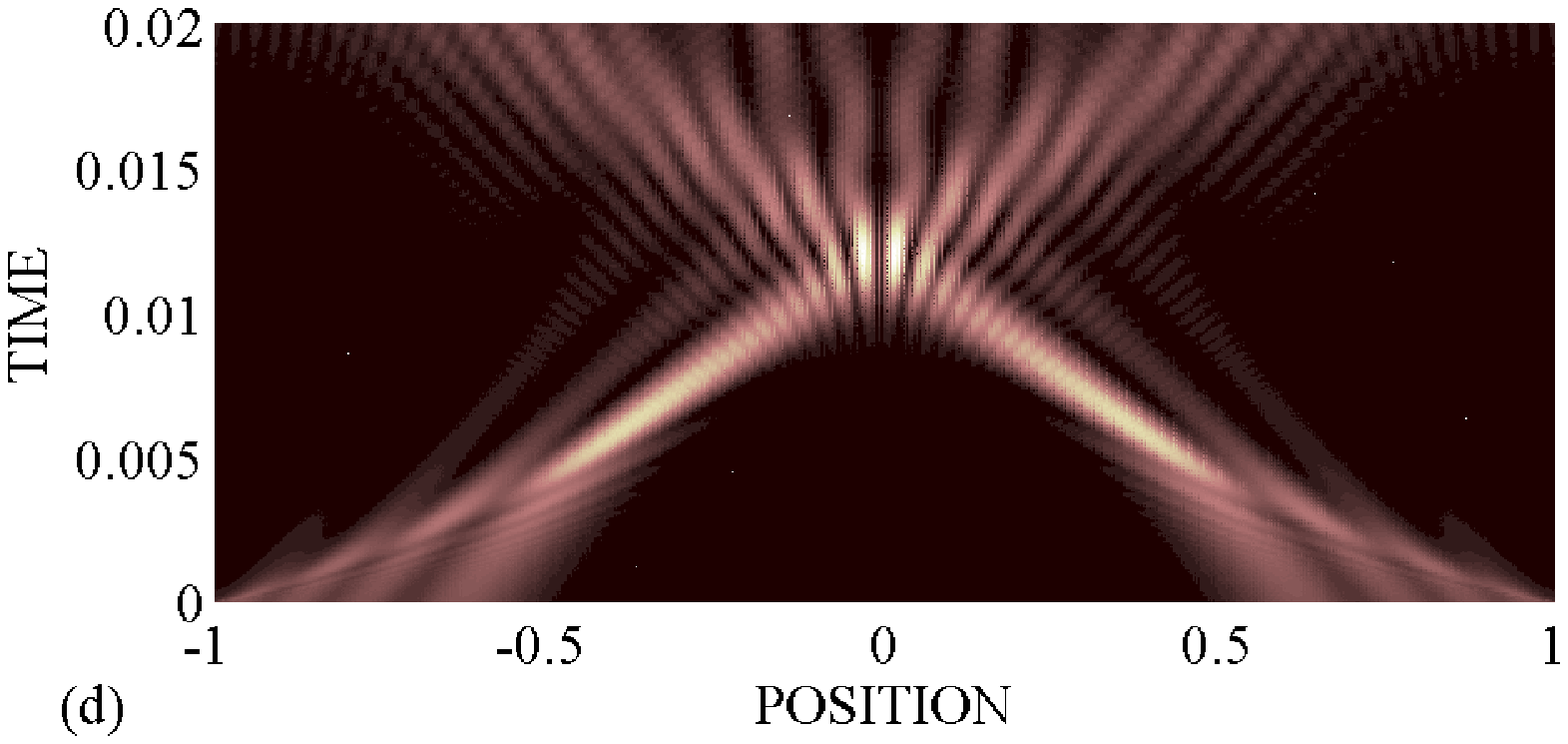}
\caption{(Color Online) Quantum carpets generated by $\varphi_{n,\pi/2,+}^{s,1,o}(q,t)$ corresponding to the eigenvalue closest to $t=0.01$ as a function of both position and time, for $l=1$ and $s=0$ (a), $s=5$ (b), $s=10$ (c) and $s=15$ (d), with $\gamma=\pi/2$. In this figure and in succeding figures, all units are in atomic units.}
\end{figure}
\begin{figure}
\includegraphics[width=0.23\textwidth,height=0.23\textwidth]{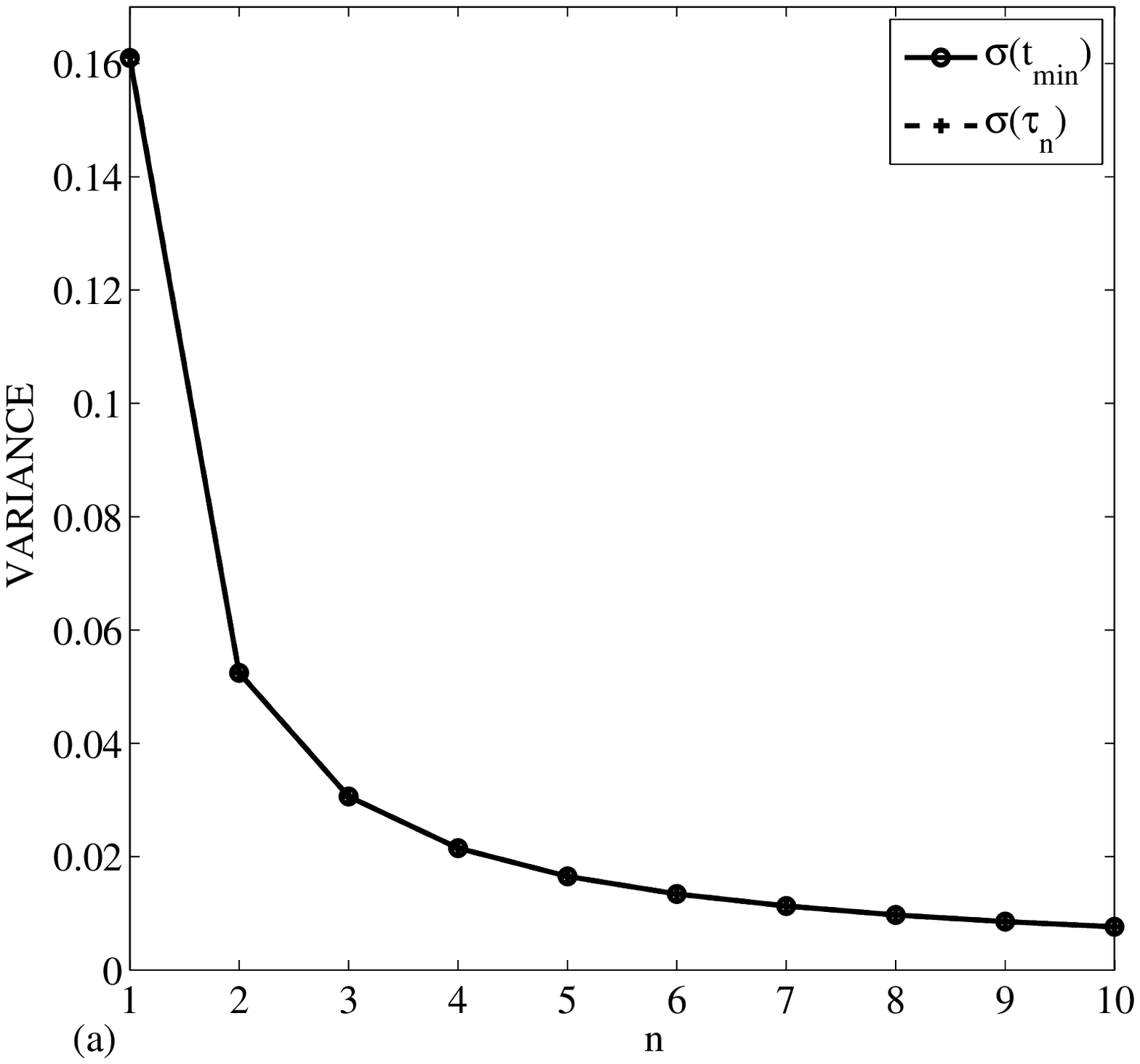}
\includegraphics[width=0.23\textwidth,height=0.23\textwidth]{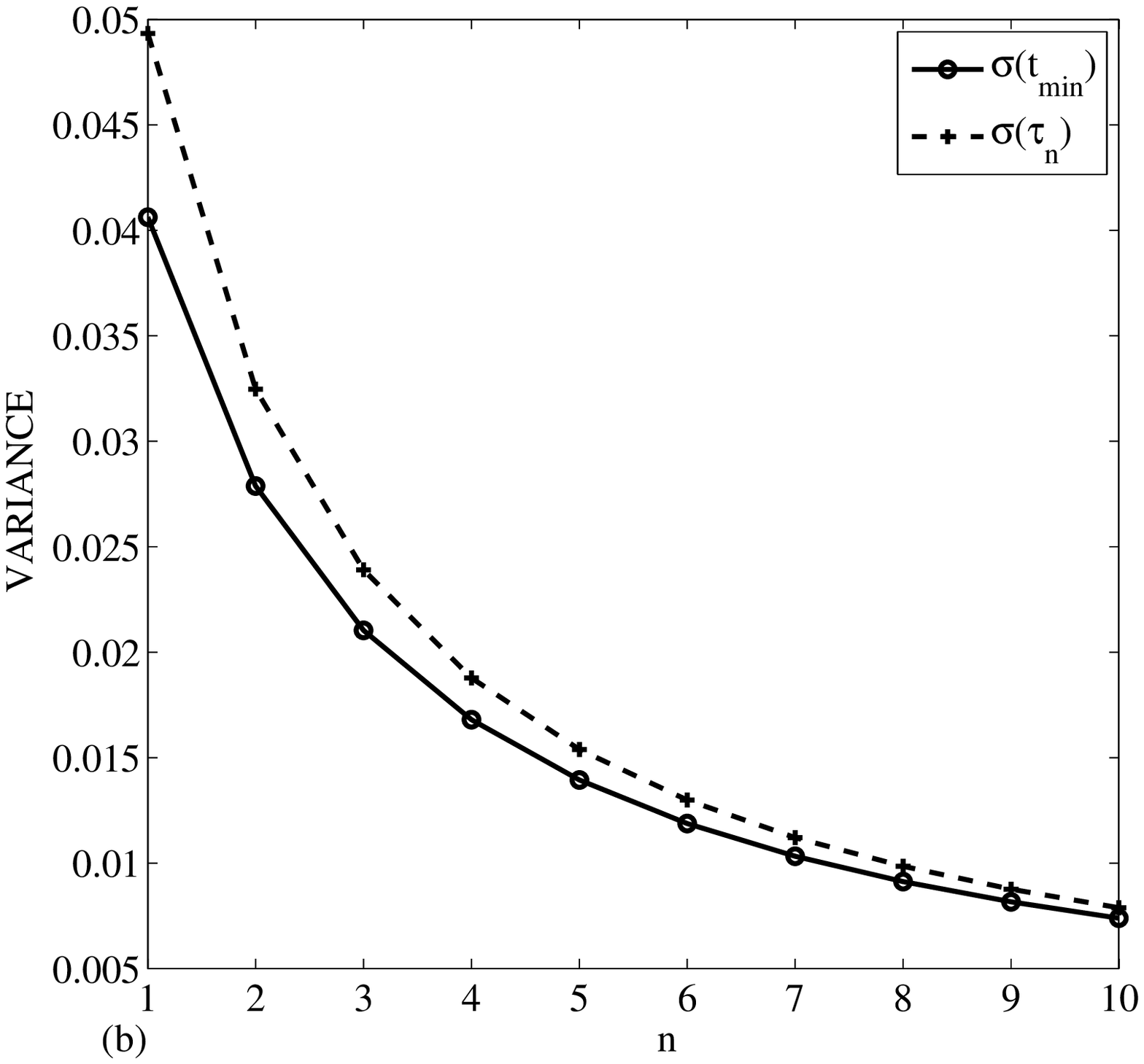}
\includegraphics[width=0.23\textwidth,height=0.23\textwidth]{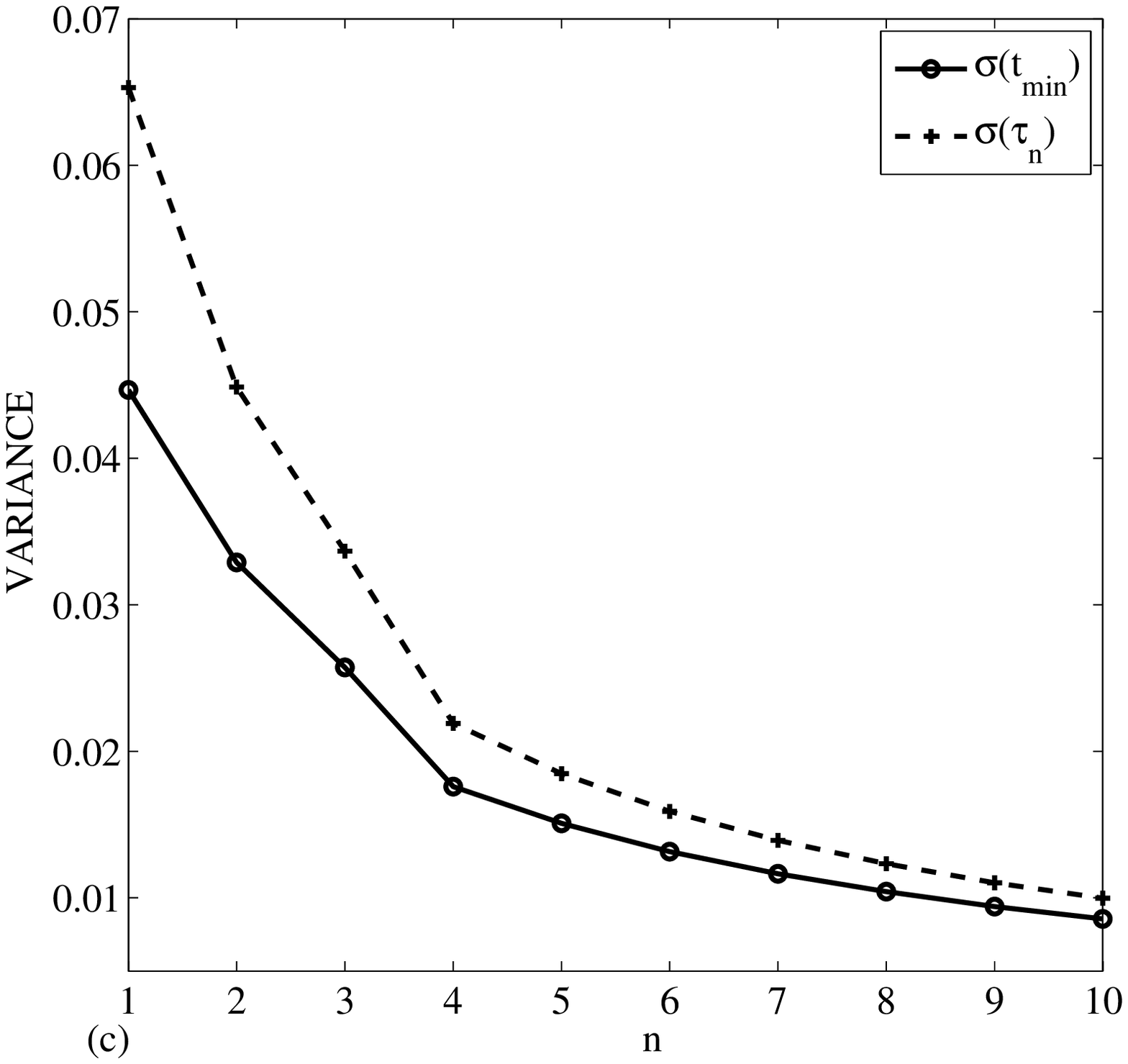}
\includegraphics[width=0.23\textwidth,height=0.23\textwidth]{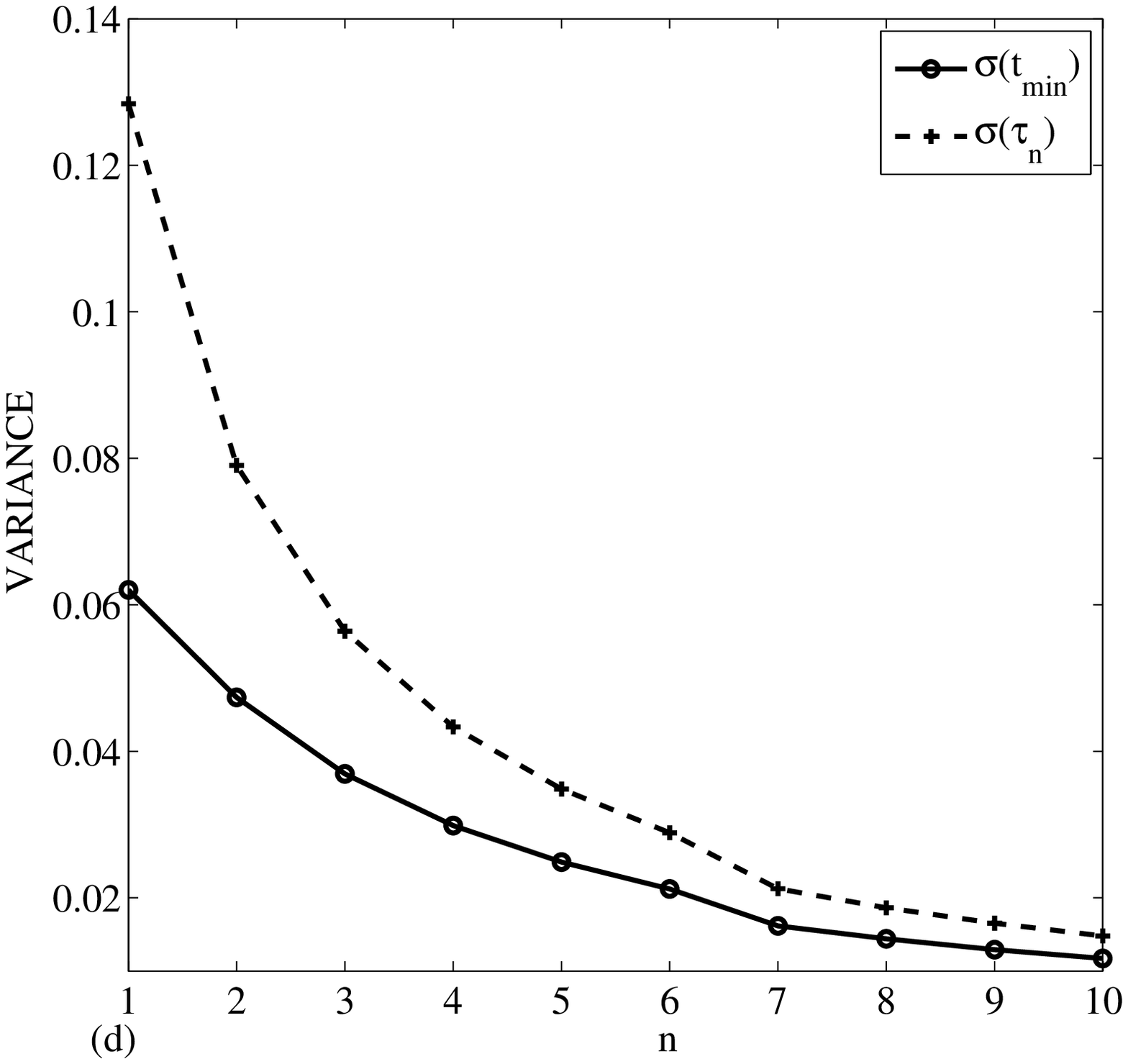}
\caption{Plots of the minimum value of the variance $\sigma^{2}(t_{min})$ of $\varphi_{n,\pi/2,+}^{s,1,o}(q,t)$ corresponding to the first ten eigenvalues as a function of eigenvalue order (solid line and circular markings) superimposed over the plots of the value of the variance at the eigenvalue $\sigma^{2}(\tau_{n})$ as a function of eigenvalue order (broken line and cross markings) for $s=0$ (a), $s=5$ (b), $s=10$ (c) and $s=15$ (d).}
\end{figure}

We first consider the time evolution of the odd parity eigenfunctions of $\opr{T}_{\pi/2,s,1}$. As shown in figure 1a, if $s=0$, $\varphi_{n,\pi/2,\nu}^{0,1,o}(q,t)$ will evolve in such a way that at $t=\tau_n$, the position probability density $\left|\varphi_{n,\pi/2,\nu}^{0,1,o}(q,t)\right|^2$ will exhibit two definite symmetric peaks with a node at the origin; this behavior leads us to call these eigenfunctions nodal eigenfunctions of $\opr{T}_{\pi/2,0,1}$. At this instant of time, the position variance of $\varphi_{n,\pi/2,\nu}^{0,1,o}(q,t)$, which we designate as $\sigma^{2}(t)$, is a minimum, as shown in figure 2a. One can then say that at $t=\pm\tau_n$, $\varphi_{n,\pi/2,\nu}^{0,1,o}(q,t)$ has unitarily arrived at the origin, by virtue of the variance attaining a minimum value.  

However, if $s\neq 0$, $\varphi_{n,\pi/2,\nu}^{s,1,o}(q,t)$ will not unitarily arrive at the origin; this is because, as shown in figures 1b to 1d, the position probability density $\left|\varphi_{n,\pi/2,\nu}^{s,1,o}(q,t)\right|^2$ will become more diffuse as $s$ increases, and the position variance of $\varphi_{n,\pi/2,\nu}^{s,1,o}(q,t)$, $\sigma^{2}(t)$, will achieve a minimum value at an instant of time not equal to the eigenvalue corresponding to $\varphi_{n,\pi/2,\nu}^{s,1,o}(q,t)$, as shown in figures 2b to 2d. Furthermore, the instant of time when $\sigma^{2}(t)$ is a minimum, $t_{min}$, will be farther away from the eigenvalue corresponding to $\varphi_{n,\pi/2,\nu}^{s,1,o}(q,t)$, $\tau_{n}$, as $s$ increases, as shown in figures 2b to 2d. Hence, we can then conclude that the physical interpretation of the $\tau$-symmetric solution to the TE-CCR $\opr{T}_{\pi/2,0,1}$ is different from the physical interpretation of the non $\tau$-symmetric solution to the TE-CCR $\opr{T}_{\pi/2,s,1}$.
\begin{figure}
\includegraphics[width=0.23\textwidth,height=0.23\textwidth]{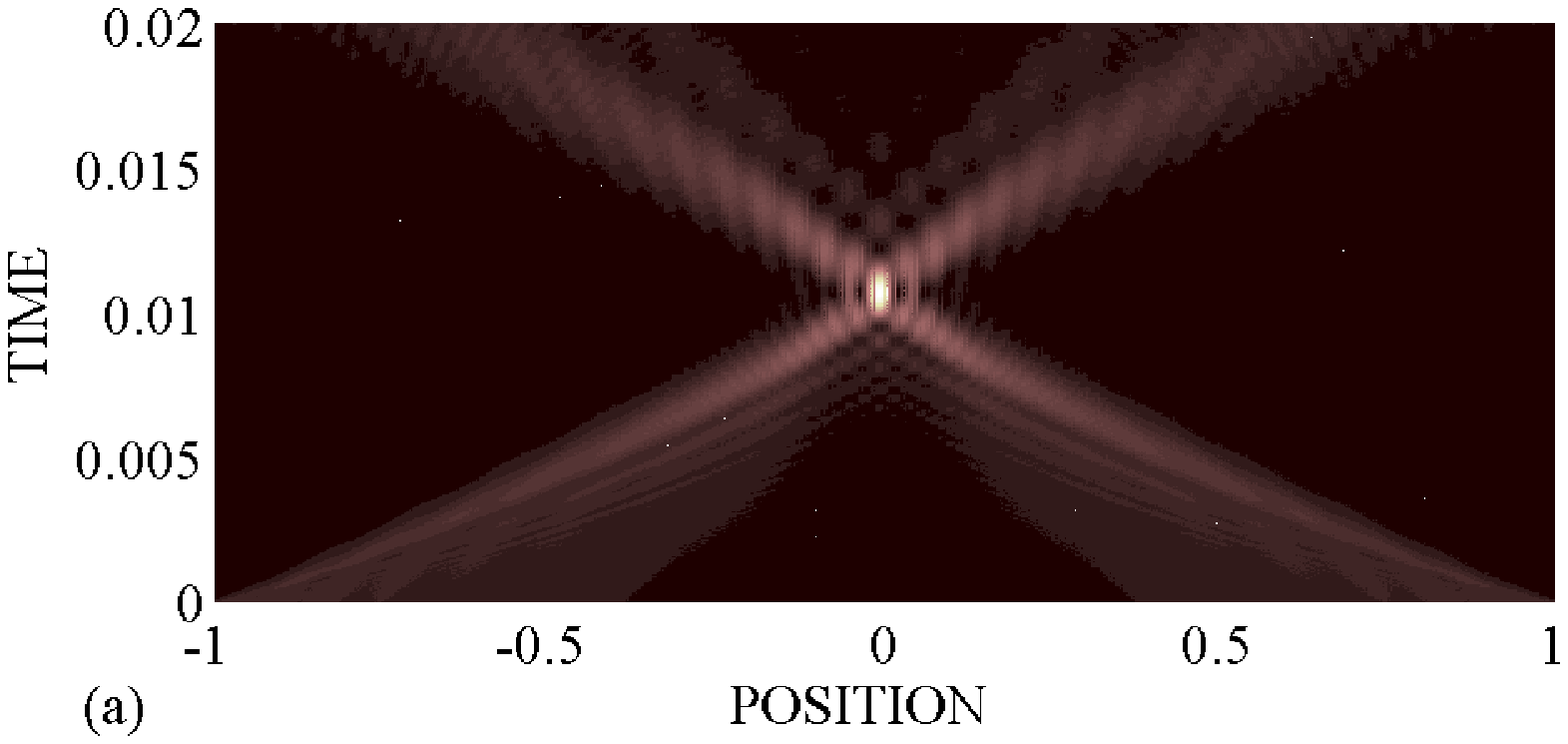}
\includegraphics[width=0.23\textwidth,height=0.23\textwidth]{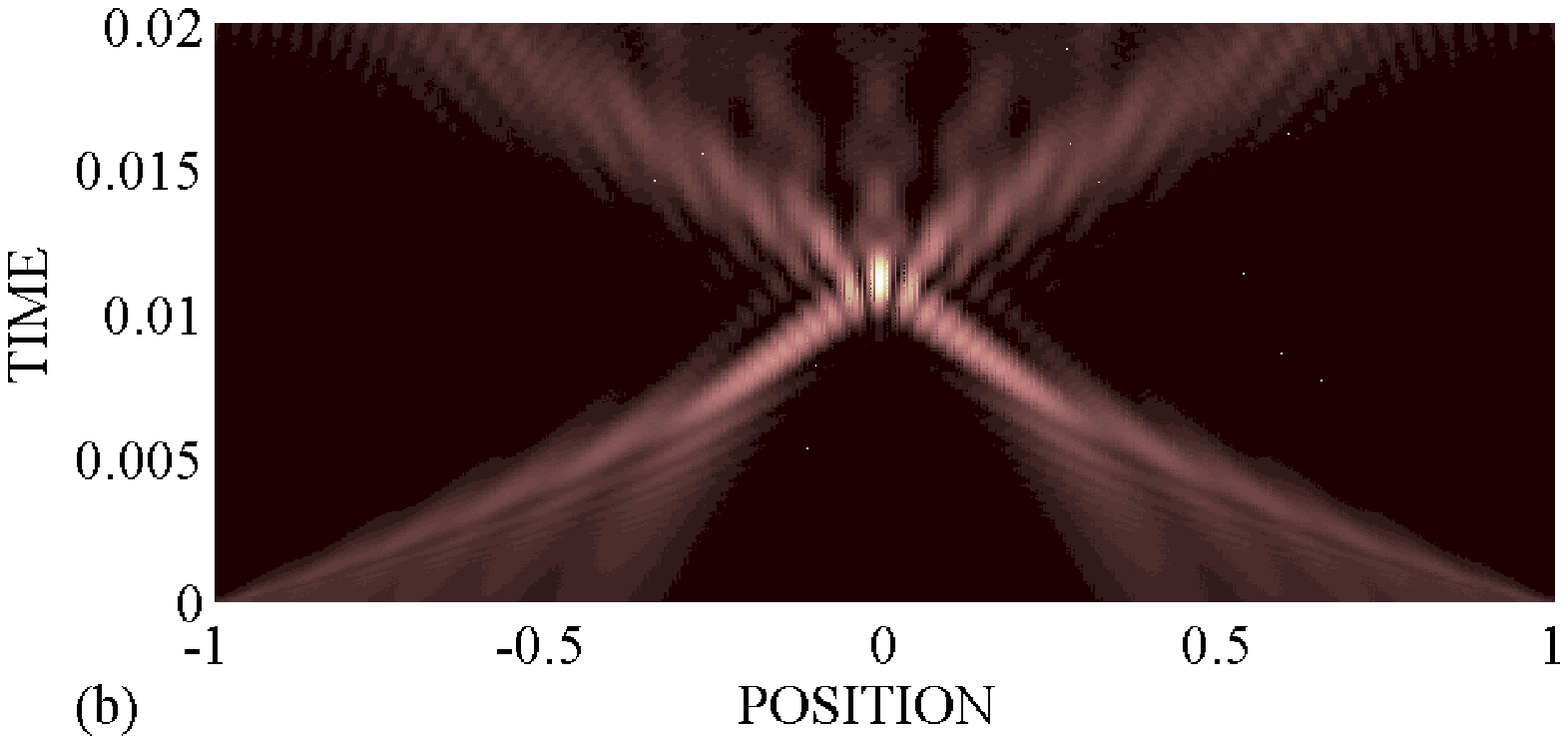}
\includegraphics[width=0.23\textwidth,height=0.23\textwidth]{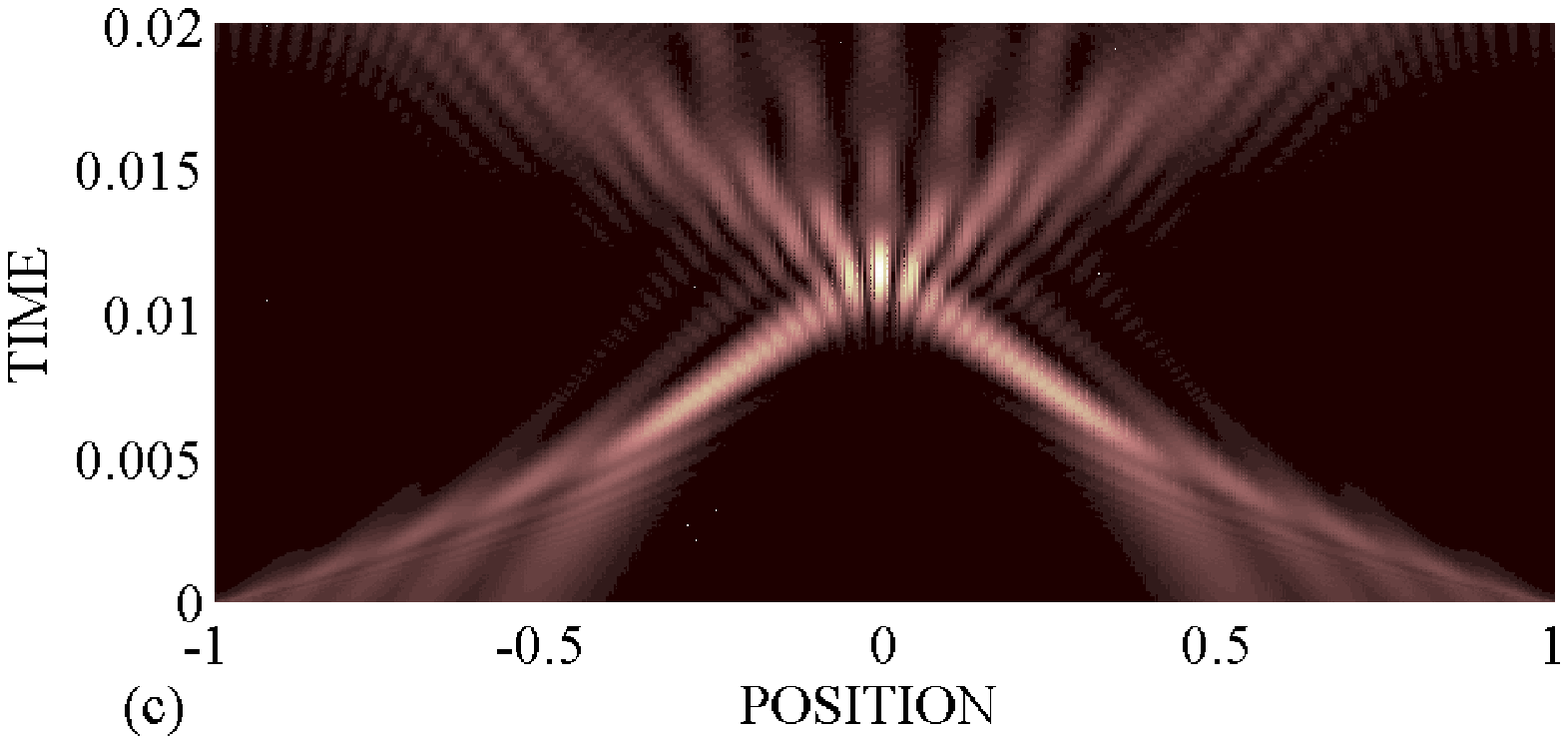}
\includegraphics[width=0.23\textwidth,height=0.23\textwidth]{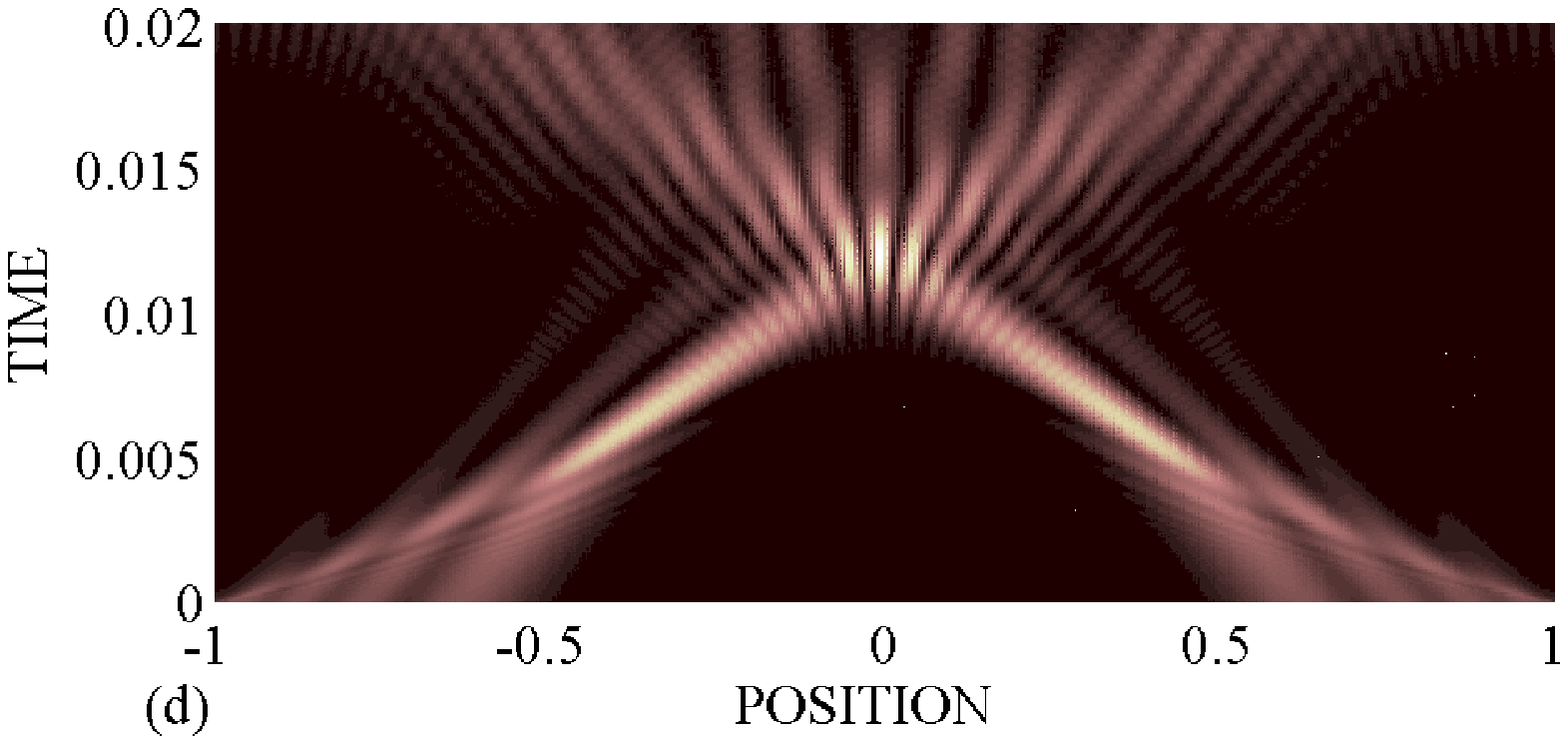}
\caption{(Color Online) Quantum carpets generated by $\varphi_{n,0,+}^{s,1,e}(q,t)$ corresponding to the eigenvalue closest to $t=0.01$ as a function of both position and time, for $l=1$ and $s=0$ (a), $s=5$ (b), $s=10$ (c) and $s=15$ (d).}
\end{figure}
\begin{figure}
\includegraphics[width=0.23\textwidth,height=0.23\textwidth]{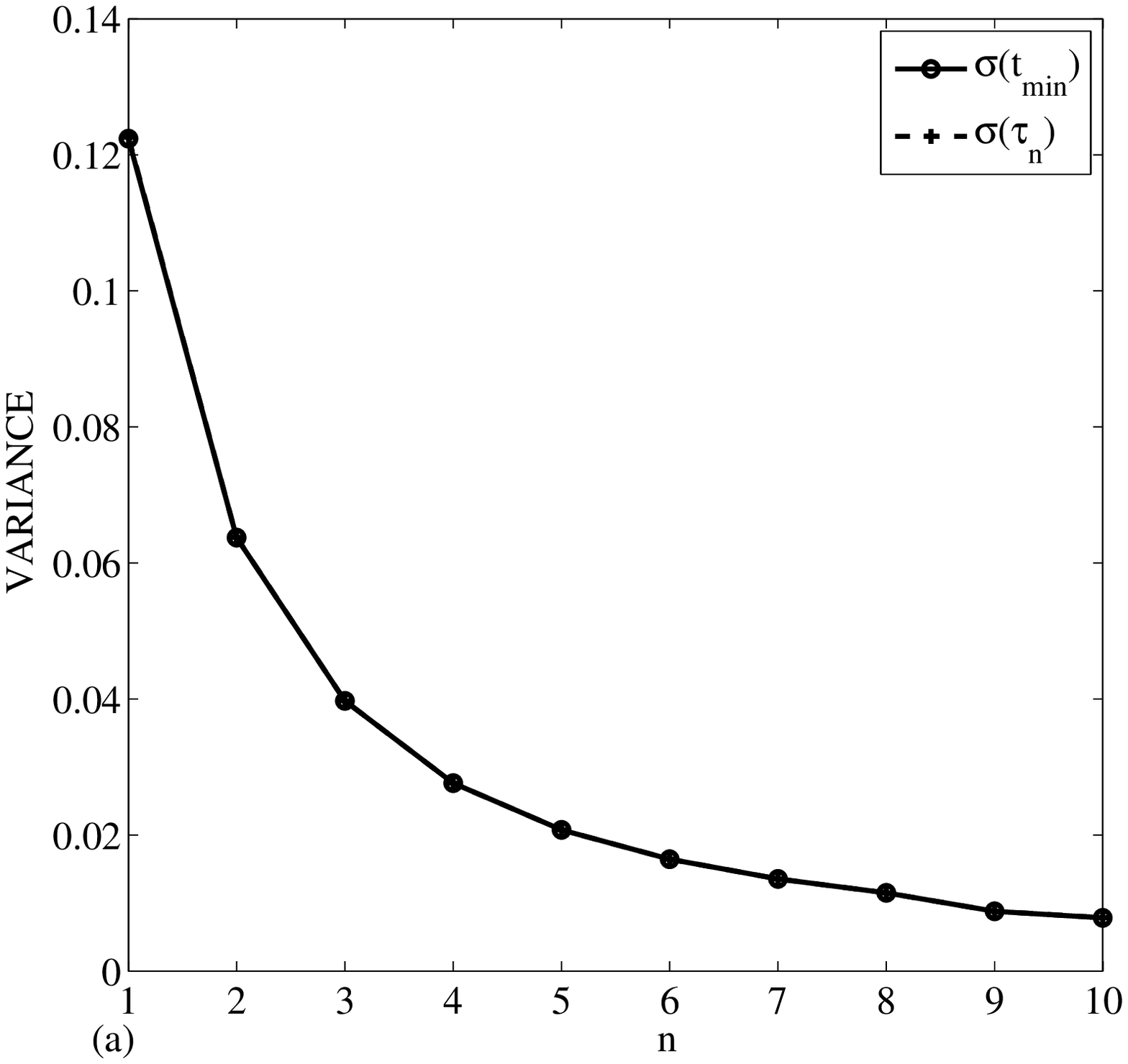}
\includegraphics[width=0.23\textwidth,height=0.23\textwidth]{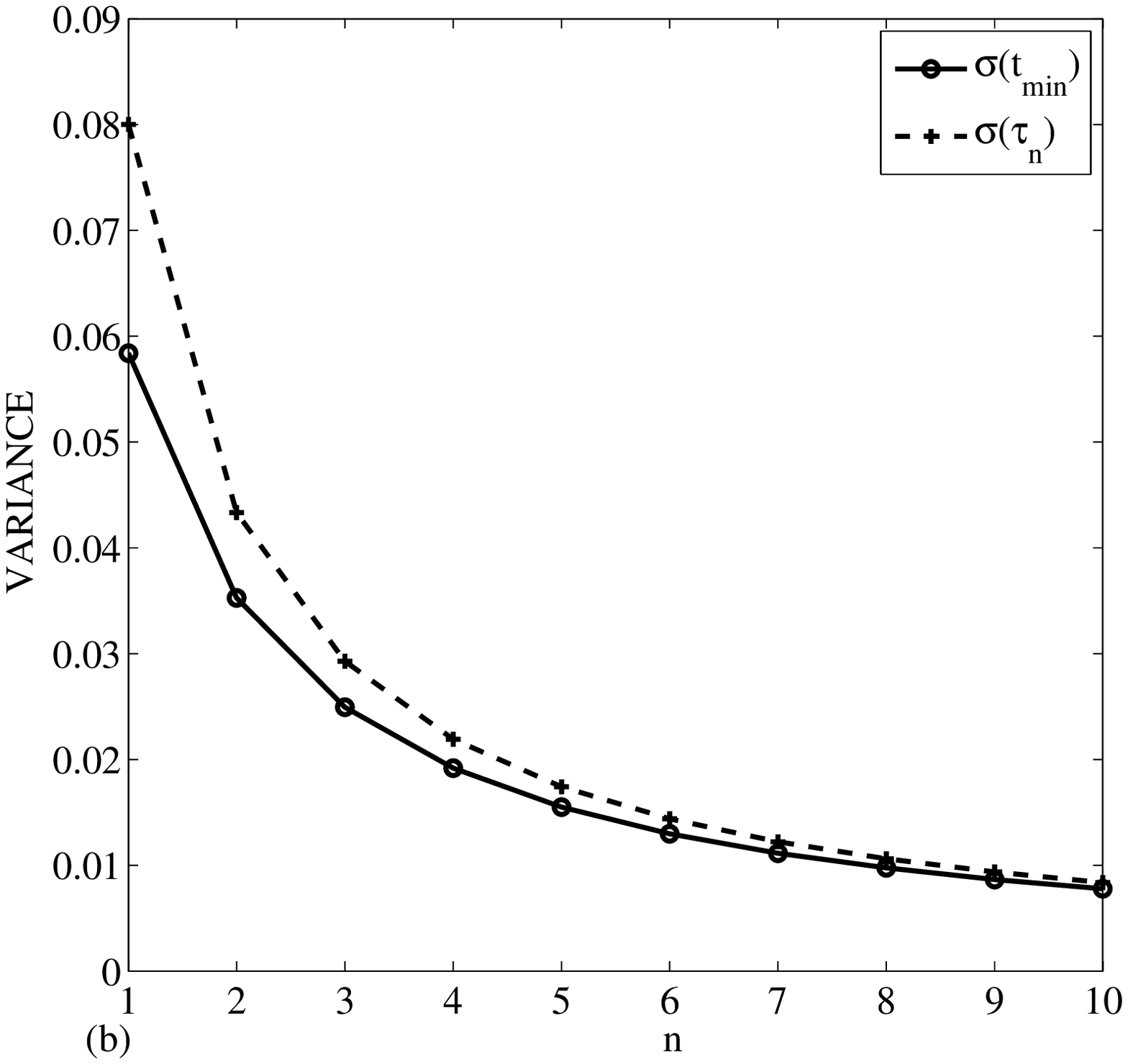}
\includegraphics[width=0.23\textwidth,height=0.23\textwidth]{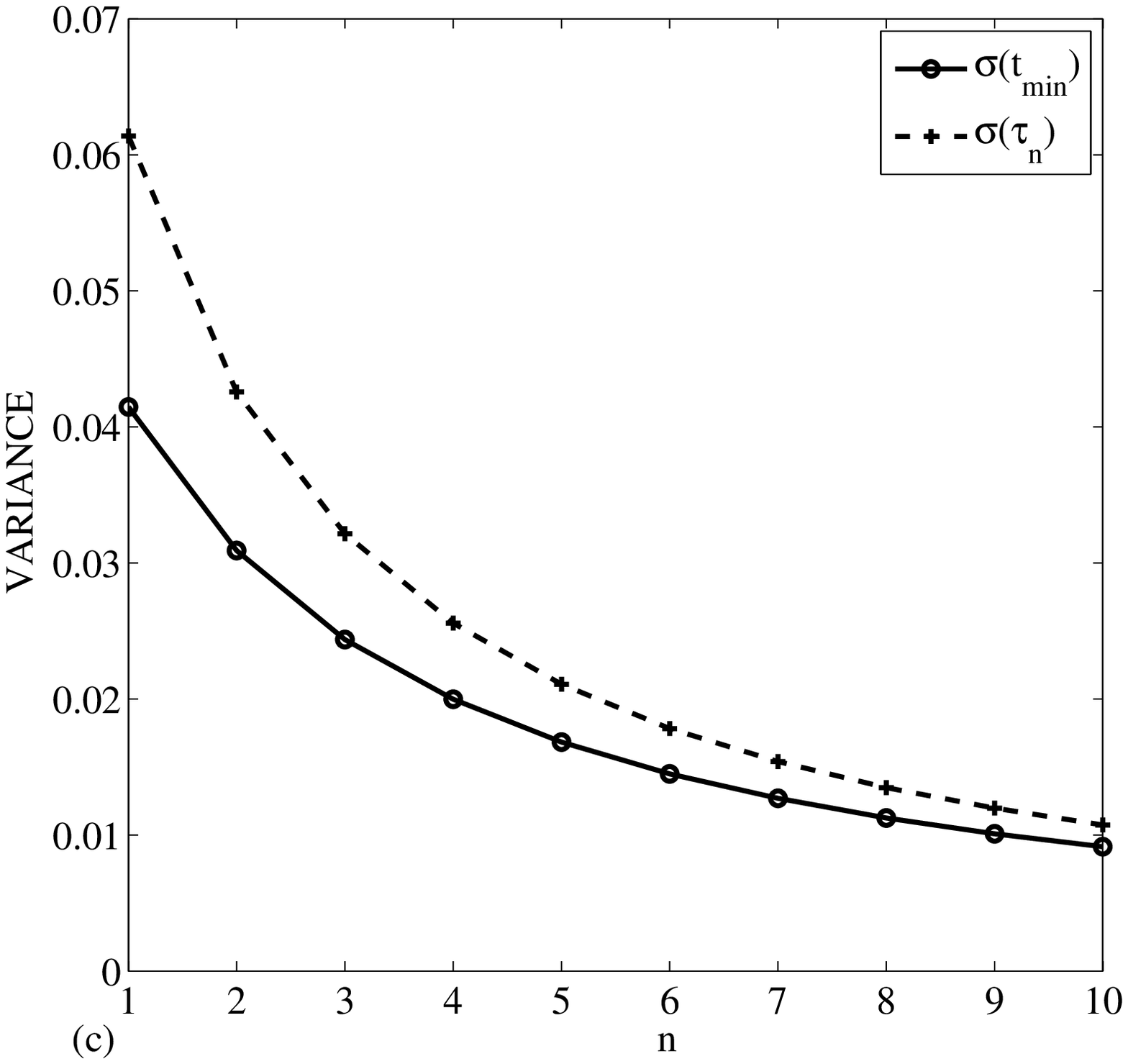}
\includegraphics[width=0.23\textwidth,height=0.23\textwidth]{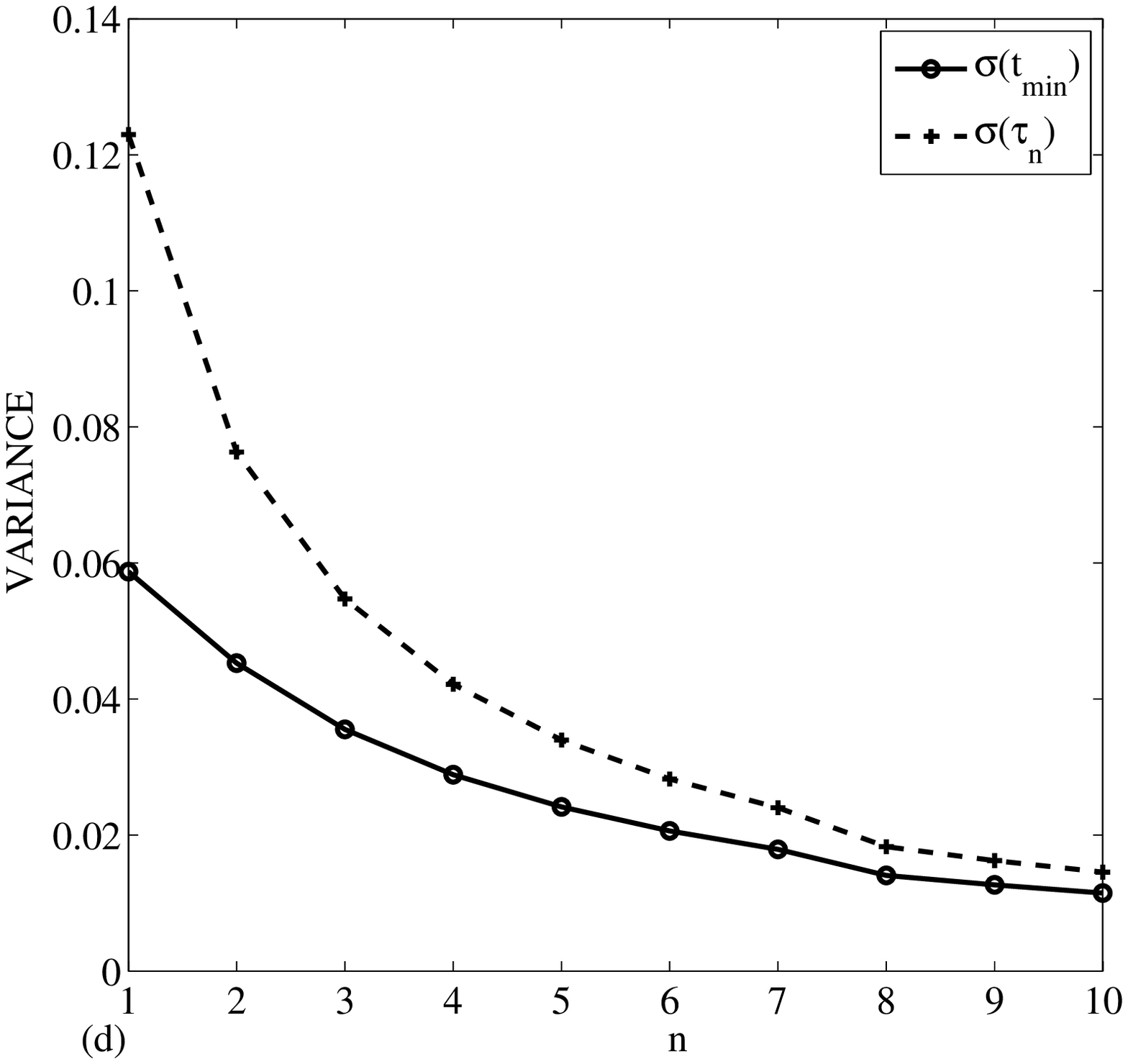}
\caption{Plots of the minimum value of the variance $\sigma^{2}(t_{min})$ of $\varphi_{n,0,+}^{s,1,e}(q,t)$ corresponding to the first ten eigenvalues as a function of eigenvalue order (solid line and circular markings) superimposed over the plots of the value of the variance at the eigenvalue $\sigma^{2}(\tau_{n})$ as a function of eigenvalue order (broken line and cross markings) for $s=0$ (a), $s=5$ (b), $s=10$ (c) and $s=15$ (d).}
\end{figure}
Let us now consider the time evolution of the even parity eigenfunctions of $\opr{T}_{0,s,1}$. For the time reversal symmetric solution to the TE-CCR $\opr{T}_{0,1}$, its time evolved even parity eigenfunctions $\varphi_{n,0,\nu}^{0,1,e}(q,t)$ will evolve in such a way that at $t=\tau_n$, the position probability density $\left|\varphi_{n,0,\nu}^{0,1,e}(q,t)\right|^2$ will exhibit a single peak centered about the the origin, as shown in figure 3a; this behavior leads us to call these eigenfunctions non-nodal eigenfunctions of $\opr{T}_{0,1}$ (to distinguish them from the nodal eigenfunctions). At this instant of time, the variance of $\varphi_{n,0,\nu}^{0,1,e}(q,t)$ is a minimum, as shown in figure 4a. One can then say that at $t=\tau_n$, $\varphi_{n,0}^{0,1,o}(q,t)$ has unitarily arrived at the origin, by virtue of the variance attaining a minimum value and the position expectation value achieving the closest possible value to zero at this instant of time. 

However, if $s\neq 0$, $\varphi_{n,0,\nu}^{s,1,e}(q,t)$ will not unitarily arrive at the origin; this is because, as shown in figures 3b to 3d, the position probability density $\left|\varphi_{n,0,\nu}^{s,1,e}(q,t)\right|^2$ will become more diffuse as $s$ increases, and $\sigma^{2}(t)$ will achieve a minimum value at an instant of time not equal to the eigenvalue corresponding to $\varphi_{n,0,\nu}^{s,1,e}(q,t)$, as shown in figures 4b to 4d. Furthermore, the instant of time when $\sigma^{2}(t)$ is a minimum, $t_{min}$, will be farther away from the eigenvalue corresponding to $\varphi_{n,0,\nu}^{s,1,e}(q,t)$, $\tau_{n}$, as $s$ increases, as shown in figures 4B to 4D. Hence, we can then conclude that the physical interpretation of the time reversal symmetric solution to the TE-CCR $\opr{T}_{0,1}$ is different from the physical interpretation of the non time reversal symmetric solution to the TE-CCR $\opr{T}_{0,s,1}$.
\section{Existence of Multiple Solutions to the TE-CCR of Dense Category}
Having shown that multiple solutions to the TE-CCR of closed category can be constructed, and having shown that it is possible to physically distinguish between these multiple solutions to the TE-CCR of closed category from each other via $\tau$-symmetry, we proceed to determine whether the same is true for multiple solutions to the TE-CCR of dense category. 
\subsection{Solutions to the TE-CCR of Dense Category}
Let us now consider the second category of solutions $\opr{T}_{\gamma,\alpha,2}$ to the TE-CCR. Falling under this category is the characteristic time operator, or CTO, which was first introduced by Galapon in reference \cite{gal022} and whose physical interpretation was given in reference \cite{caballar}. Now it has been shown by Galapon in reference \cite{gal022} that the CTO can generate an infinite number of self-adjoint and compact solutions to the TE-CCR of dense category, by adding a term which commutes with the system's Hamiltonian $\opr{H}_{\gamma}$. The resulting time operator, which we call the generalized characteristic time operator (GTO), will have the following form in position representation:
\begin{equation}
\left(\opr{T}_{\gamma,\alpha,2}\varphi\right)(q)=\int_{-l}^{l}\left\langle q\right|\opr{T}_{\gamma,\alpha,2}\left|q'\right\rangle\varphi(q')dq'
\label{cto}
\end{equation}
where the kernel $\left\langle q\right|\opr{T}_{\gamma,\alpha,2}\left|q'\right\rangle$ of the GTO $\opr{T}_{\gamma,s,2}$ has the following explicit form:
\begin{eqnarray}
&&\left\langle q\right|\opr{T}_{\gamma,\alpha,2}\left|q'\right\rangle=i\hbar\sum_{k,k'=-\infty}^{+\infty}\!'\frac{\phi_{k}^{\gamma}(q)\phi_{k'}^{\gamma*}(q')}{E_{k}-E_{k'}}+\nonumber\\
&&\sum_{k=-\infty}^{\infty}\alpha_{k}\phi_{k}^{\gamma}(q)\phi_{k}^{\gamma*}(q')
\label{gtokern}
\end{eqnarray}
where the prime in the summation denotes that $k\neq k'$ and $\alpha_k$ is an element of a bounded sequence of real numbers $\alpha=\left\{\alpha_{k}\right\}_{k=1}^{\infty}$, which we call the $\alpha$ sequence corresponding to the GTO. Note that the first term of equation \ref{gtokern} is actually the kernel for the CTO, while the second term of equation \ref{gtokern} can be shown to commute with the Hamiltonian of the system. As such, this implies that the GTO is still a solution to the TE-CCR of dense category, with the canonical domain of the GTO identical to the canonical domain of the CTO. Furthermore, imposing appropriate conditions on the elements of the real valued sequence $\alpha$, e. g. $\sum_{k=-\infty}^{\infty}\left|\alpha_{k}\right|^{2}<\infty$ ensures that equation \ref{gtokern} is square integrable, which means the GTO is self-adjoint and compact. Because there are an infinite number of $\alpha$ sequences that can be chosen to ensure that the GTO is self-adjoint, compact and a solution to the TE-CCR of dense category, there now arises the question of whether we can mathematically and physically distinguish between multiple GTOs without having to resort to determining the explicit form of their corresponding $\alpha$ sequences. We will answer this question in the next portion of the paper, but before we do so, let us first examine the physical interpretation of the CTO and how it can be shown to be mathematically distinct from other forms of the GTO.
\subsection{Physical interpretation of the CTO and how it is mathematically distinct from the GTO}
The physical interpretation of the CTO was first stated in reference \cite{caballar}, and was made using the function
\begin{equation}
P_{t}[n,n']=\left|\left\langle\varphi_{n',\gamma,\nu}\right|U_{t}\left|\varphi_{n,\gamma,\nu}\right\rangle\right|^{2}
\label{transprob}
\end{equation}
which is also known as the transition probability, or the probability that the time evolved CTO eigenfunctions $\ket{\varphi_{n,\gamma,\nu}(t)}$ will make a transition to $\ket{\varphi_{n',\gamma,\nu}}$ at an instant of time $t$. As shown in figure 5a, the plot of $P_{t}[n,n']$ as a function of time will have a single, well-defined peak whose maximum value is numerically close to, if not equal to, one, at an instant of time $t_{max}$; furthermore, the peak will be very narrow, signifying that any measurement of the instant when $\ket{\varphi_{n,\gamma,\nu}}$ will make a transition to $\ket{\varphi_{n',\gamma,\nu}}$ will be very accurate. On the other hand, as shown in figure 6a, plotting $t_{max}$ as a function of $\tau_{n}-\tau_{n'}$ will result in a linear graph whose slope is numerically very close to, if not equal to, one, with $\tau_{n}-\tau_{n'}$ asymptotically approaching zero, signifying that $t_{max}$ is numerically very close to, if not equal to, $\tau_{n}-\tau_{n'}$. Based on these plots, the CTO can be interpreted as a time operator whose time-evolved eigenfunctions $\ket{\varphi_{n,\gamma,\nu}(t)}=U_{t}\ket{\varphi_{n,\gamma,\nu}}$ will make transitions to other CTO eigenfunctions $\ket{\varphi_{n',\gamma,\nu}}$ at an instant of time very close to, if not equal to, $t=\tau_{n}-\tau_{n'}$, the difference between the eigenvalues corresponding to those eigenfunctions, so long as the difference between those eigenvalues asymptotically approaches zero.

Now the CTO and the GTO are distinct from each other with respect to $\tau$-symmetry. In particular, the CTO is $\tau$-symmetric; this can be shown by making use of the observation that $E_{k,\gamma}=E_{-k,-\gamma}$. However, it can be shown that the GTO $\opr{T}_{\gamma,\alpha,2}$ does not satisfy $\tau$-symmetry; this is due to the presence of the $\alpha$ sequence in the GTO's kernel. Hence, we can distinguish between the CTO and the GTO via $\tau$-symmetry, with the CTO $\opr{T}_{\gamma,2}$ being $\tau$-symmetric and the GTO $\opr{T}_{\gamma,\alpha,2}$ not $\tau$-symmetric. Also, the $\alpha$ sequence can then be seen as a $\tau$-symmetry breaking sequence for a GTO so long as it has nonzero elements.
\subsection{Physical Implications of $\tau$-Symmetry on the GTO}
We now investigate the effects of the presence or absence of $\tau$-symmetry in the GTO. To do so, we first compute for the eignfunctions and eigenvalues of a particular form of the GTO. We then evolve the eigenfunctions of this GTO, over time, and compare the dynamics of this GTO with the dynamics of the time-evolved CTO eigenfunctions.
\subsubsection{Computation of the Eigenvalues and the Eigenfunctions of the GTO}
\begin{figure}
\includegraphics[width=0.23\textwidth,height=0.23\textwidth]{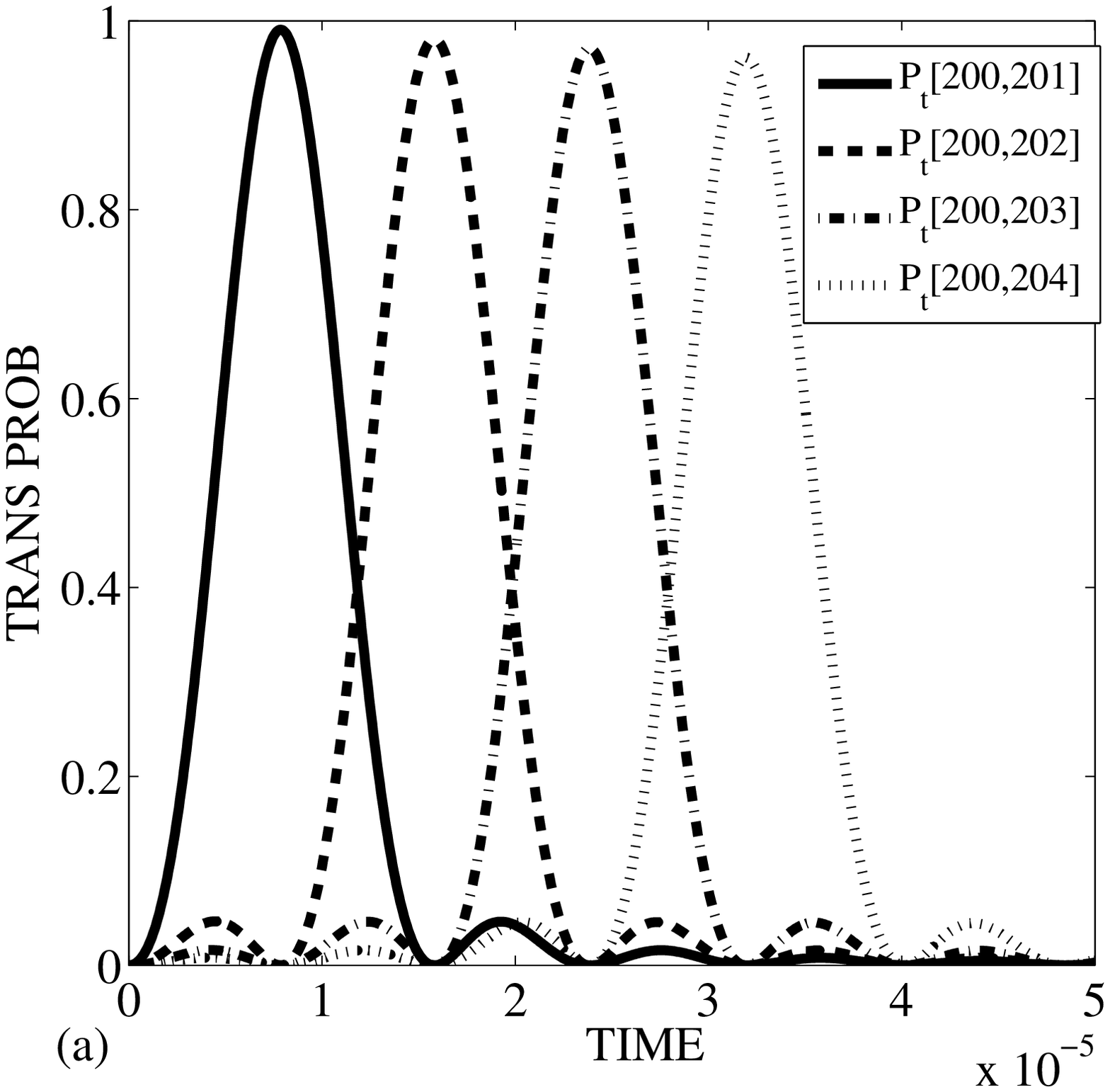}
\includegraphics[width=0.23\textwidth,height=0.23\textwidth]{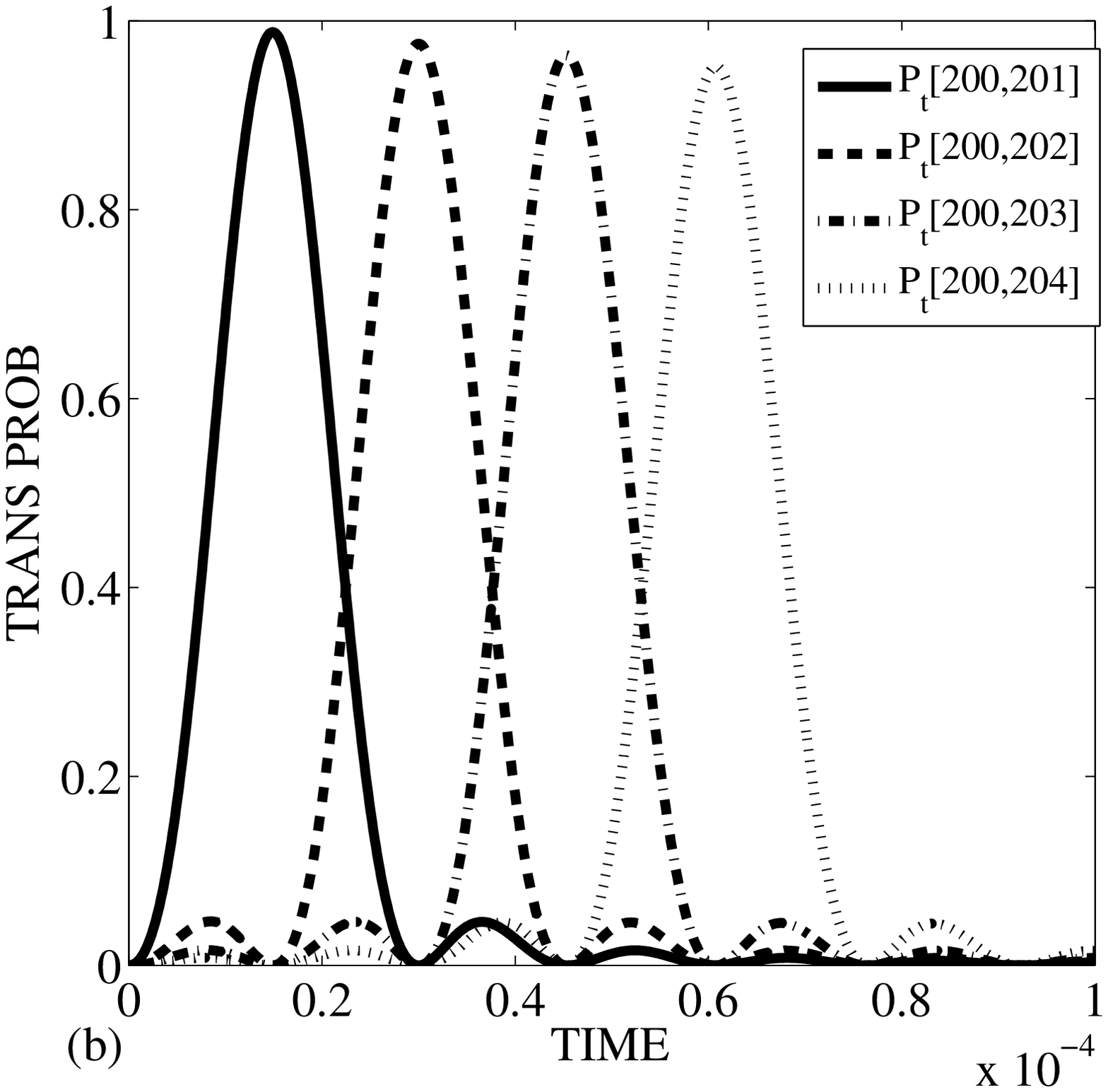}
\includegraphics[width=0.23\textwidth,height=0.23\textwidth]{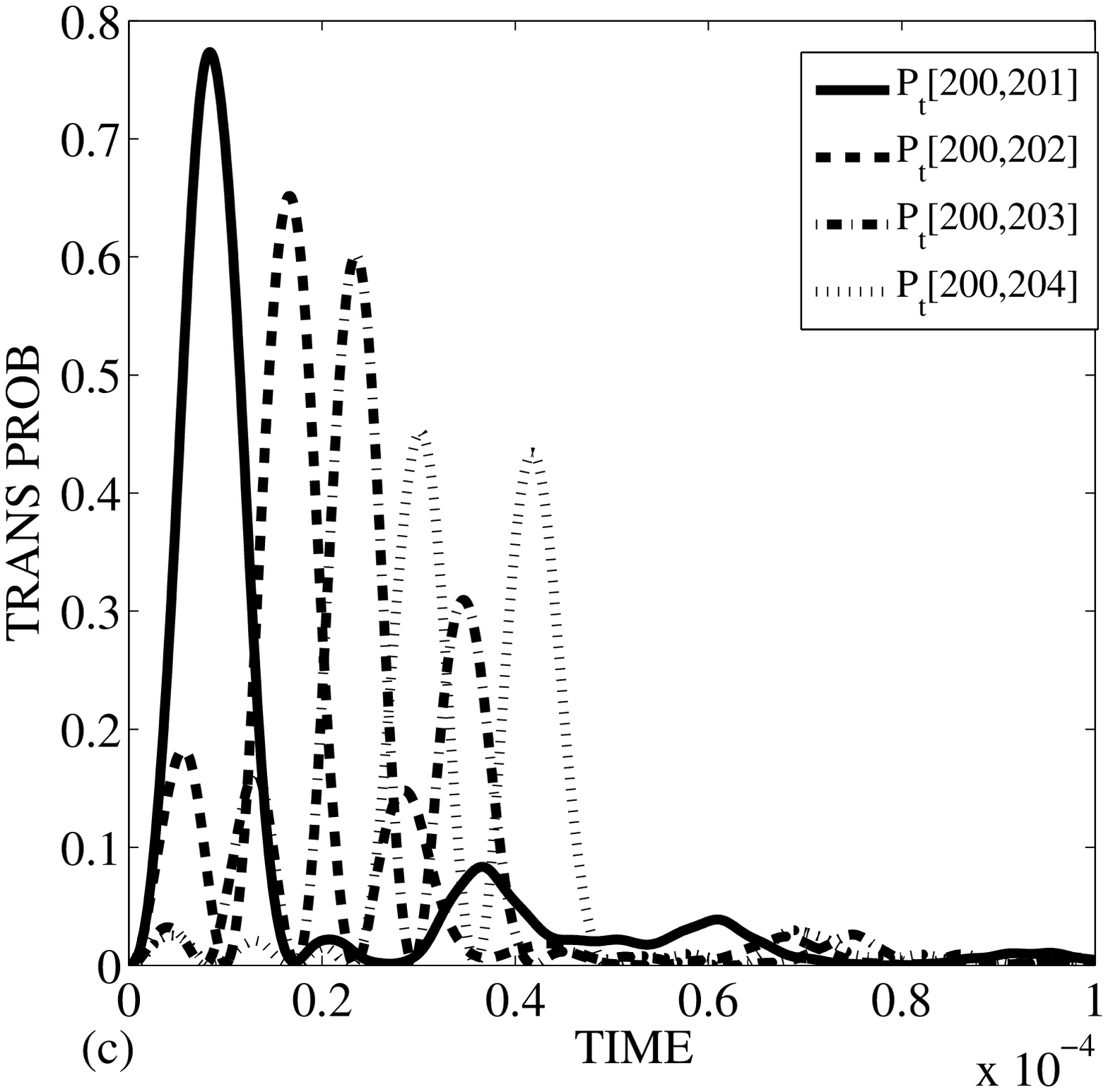}
\includegraphics[width=0.23\textwidth,height=0.23\textwidth]{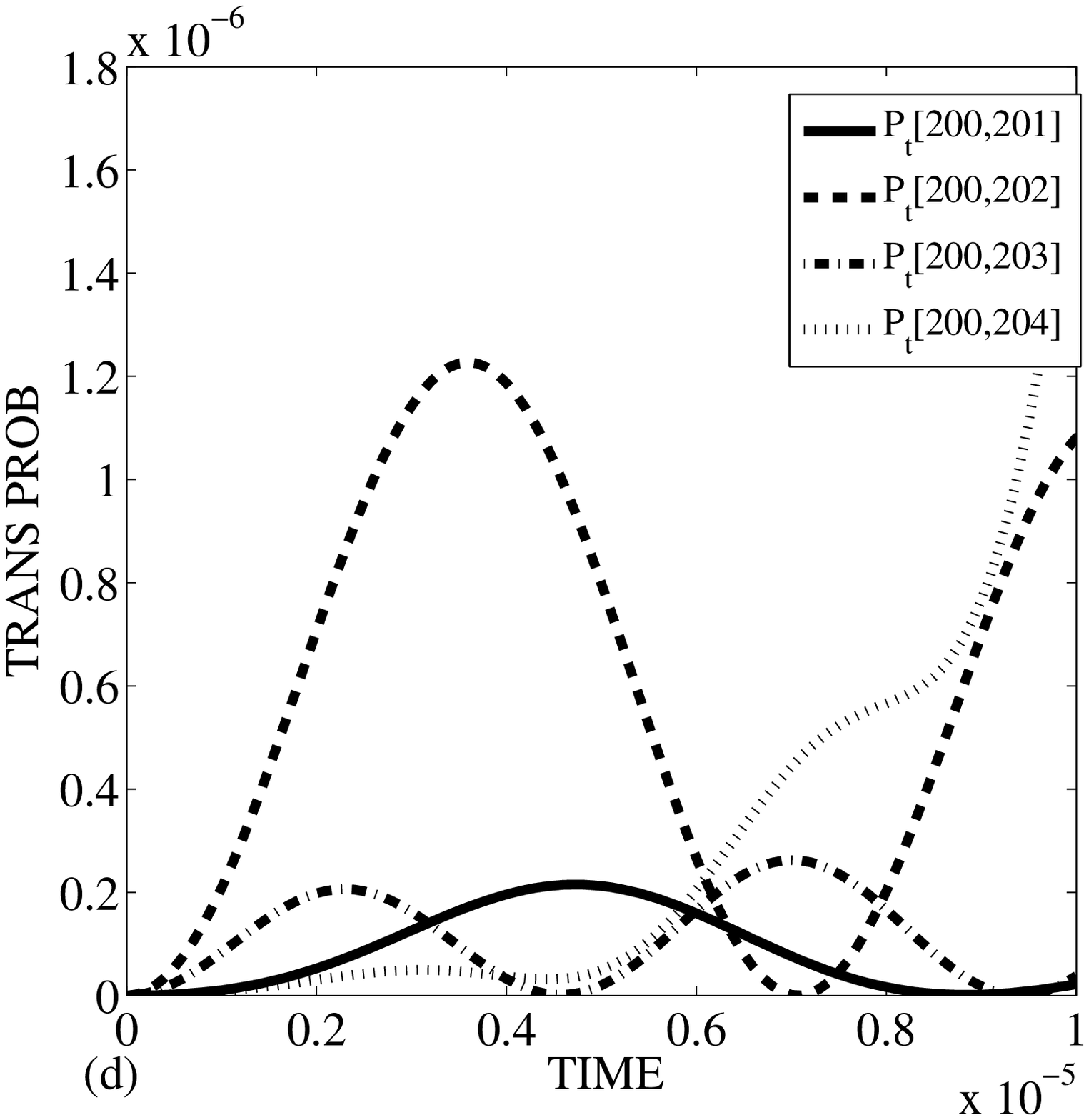}
\caption{Plot of the transition probability $P_{t}[n,n']=\left|\bra{\varphi_{n+1}}\left.U_{t}\varphi_{n}\right\rangle\right|^{2}$ corresponding to the time evolved GTO eigenfunctions as a function of time, with the GTO eigenfunctions corresponding to GTOs with $\alpha$ sequences $\alpha=\left\{0\right\}_{k=-\infty}^{\infty}$ (a), $\alpha=\left\{50E_{k,\gamma}^{-1}\right\}_{k=-\infty}^{\infty}$ (b), $\alpha=\left\{50E_{k,\gamma}^{-20}\right\}_{k=-\infty}^{\infty}$ (c) and $\alpha=\left\{50E_{k,\gamma}^{-25}\right\}_{k=-\infty}^{\infty}$ (d).}
\end{figure}
\begin{figure}
\includegraphics[width=0.23\textwidth,height=0.23\textwidth]{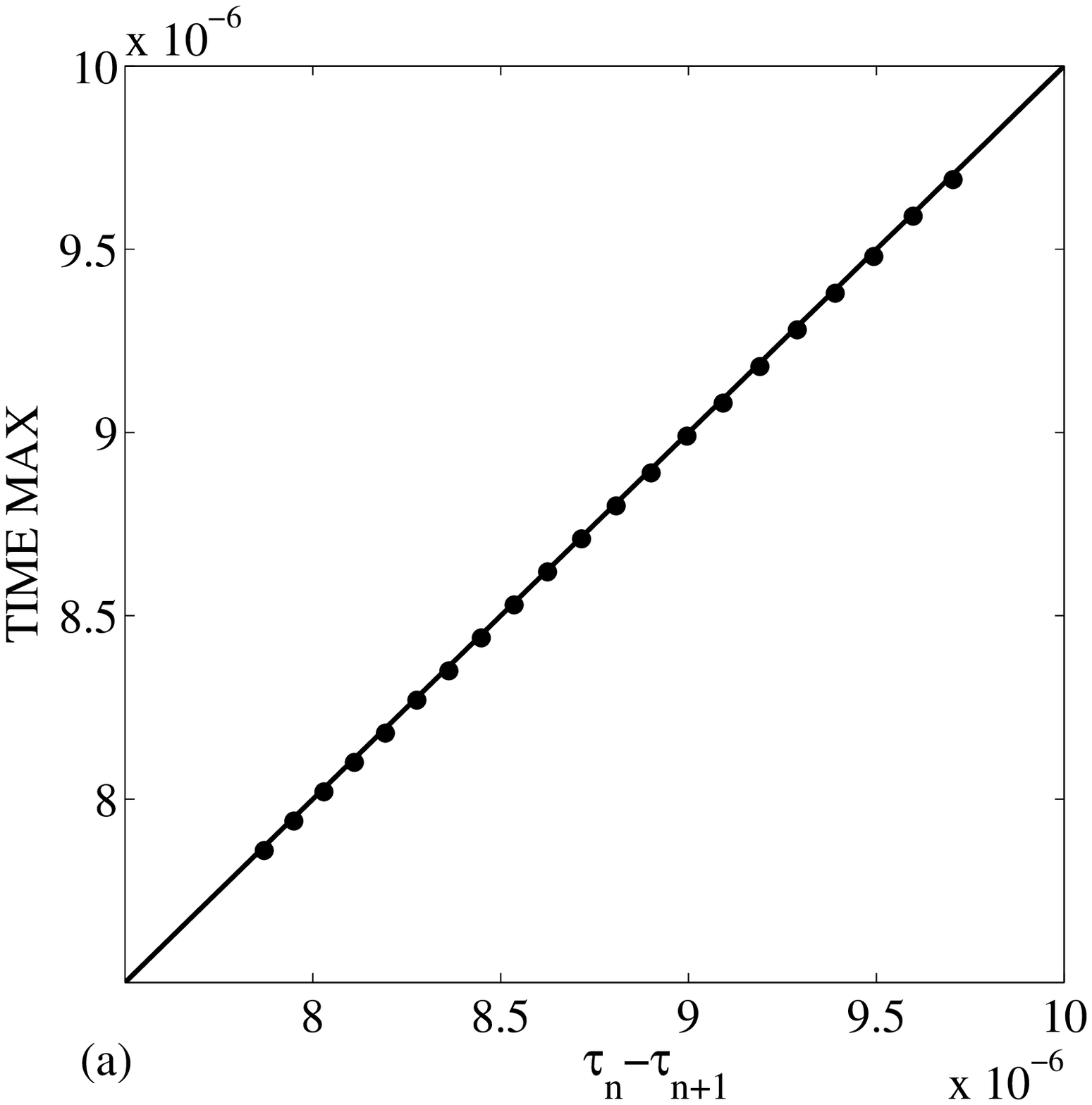}
\includegraphics[width=0.23\textwidth,height=0.23\textwidth]{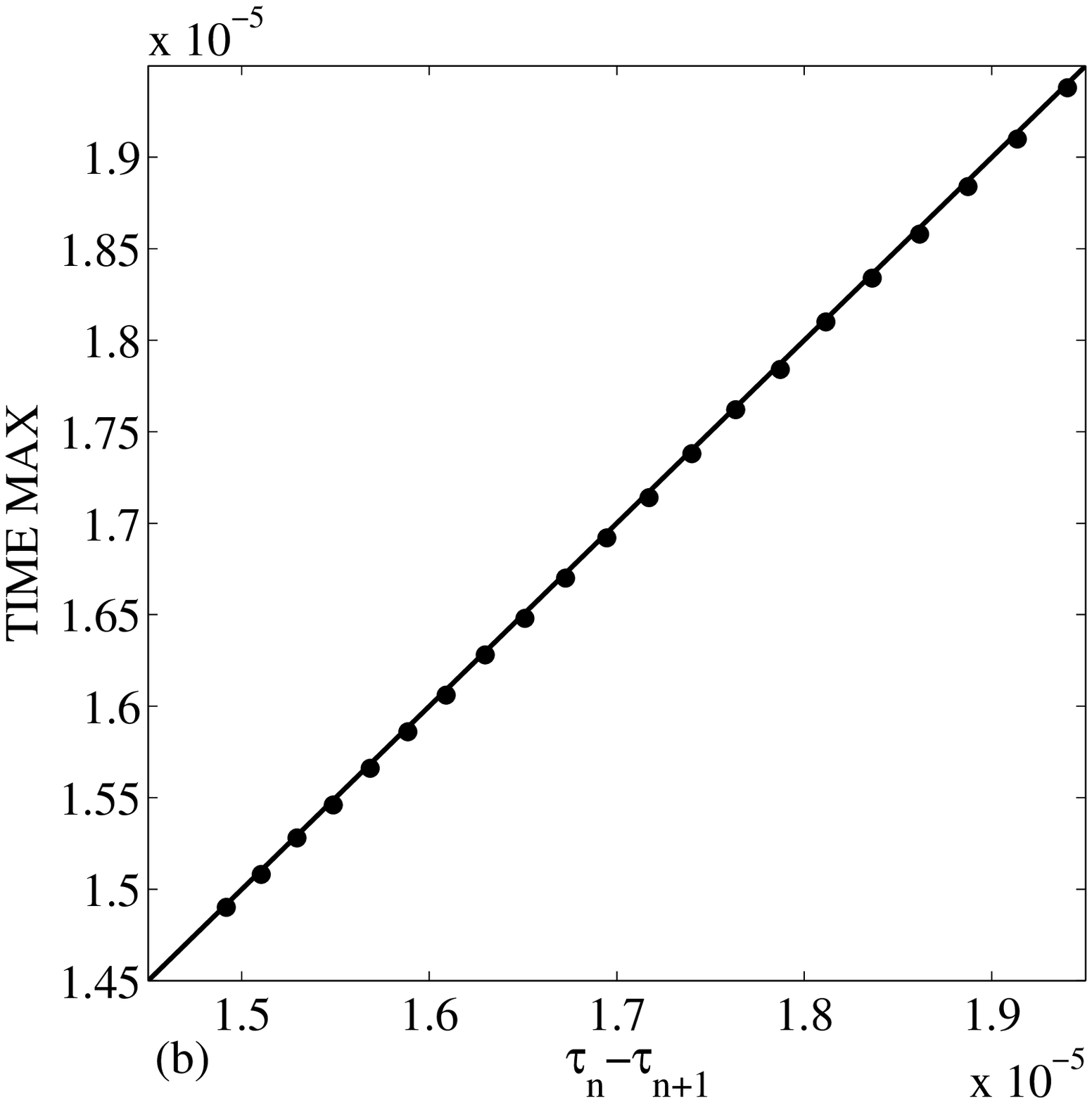}
\includegraphics[width=0.23\textwidth,height=0.23\textwidth]{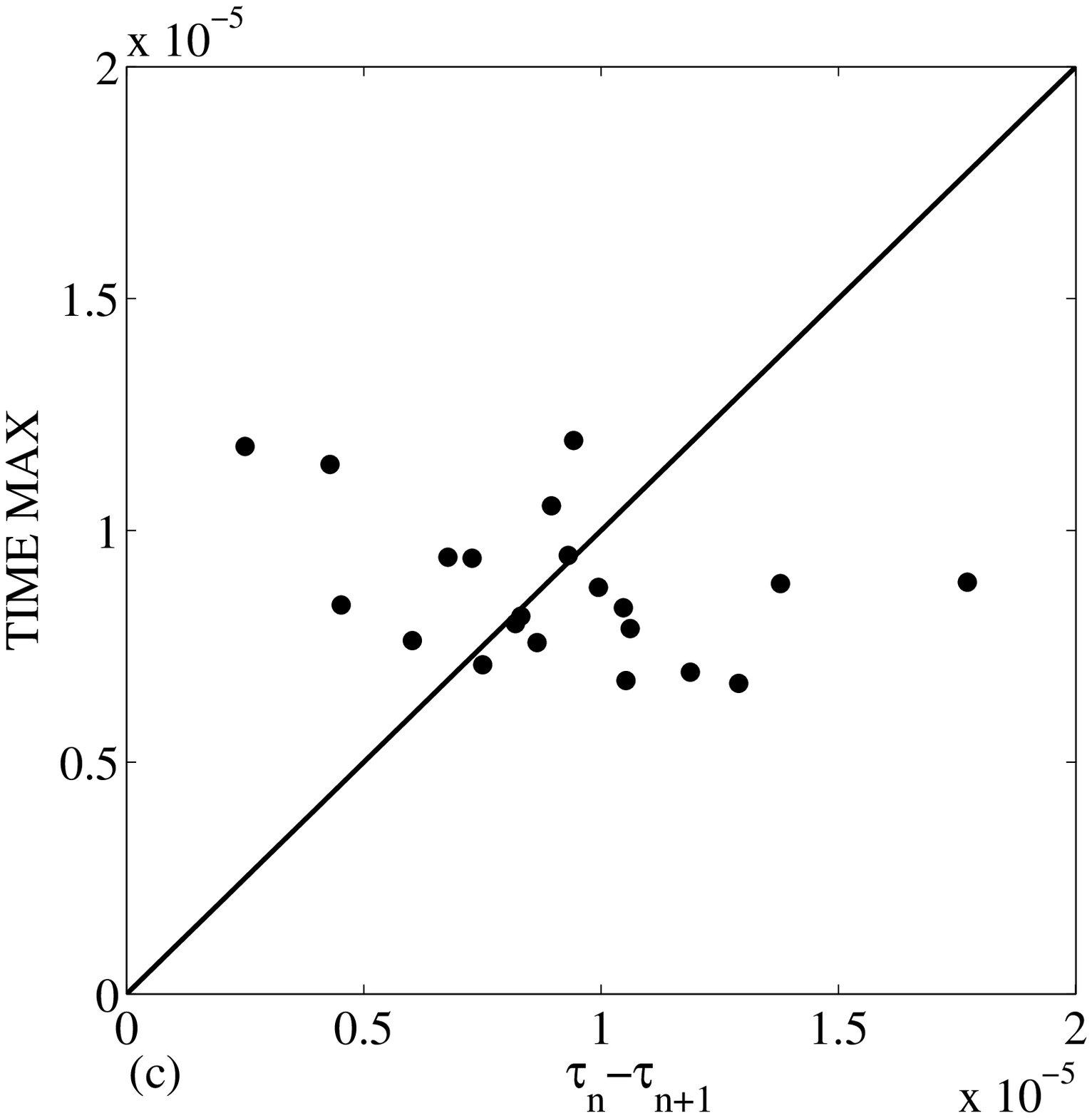}
\includegraphics[width=0.23\textwidth,height=0.23\textwidth]{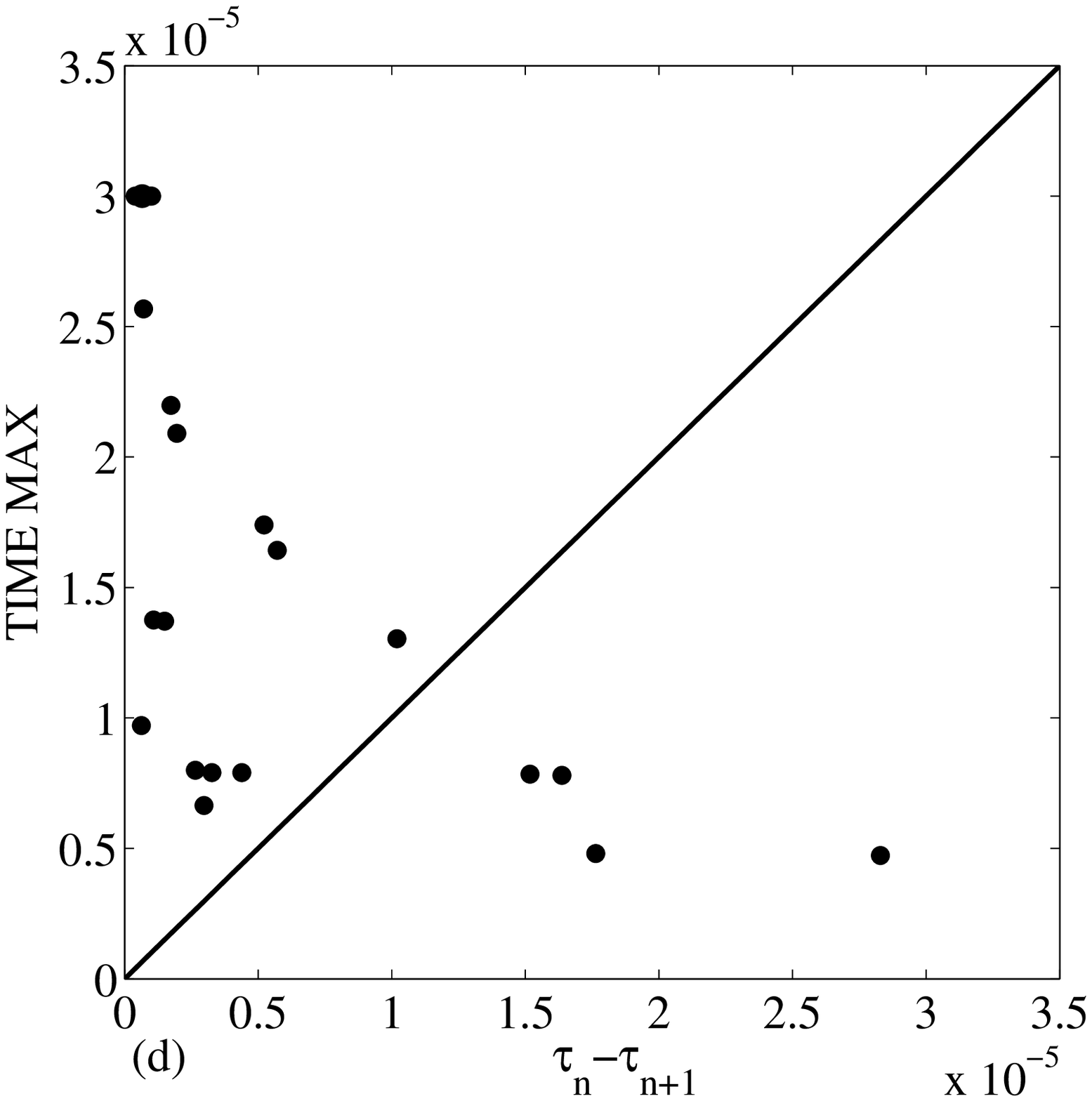}
\caption{Plot of the time when the transition probability $P_{t}[n,n+1]=\left|\bra{\varphi_{n+1}}\left.U_{t}\varphi_{n}\right\rangle\right|^{2}$ corresponding to the time evolved GTO eigenfunctions becomes a maximum as a function of the eigenvalue difference $\tau_{n}-\tau_{n+1}$, with the GTO eigenfunctions corresponding to GTOs with $\alpha$ sequences $\alpha=\left\{0\right\}_{k=-\infty}^{\infty}$ (a), $\alpha=\left\{50E_{k,\gamma}^{-1}\right\}_{k=-\infty}^{\infty}$ (b), $\alpha=\left\{50E_{k,\gamma}^{-20}\right\}_{k=-\infty}^{\infty}$ (c) and $\alpha=\left\{50E_{k,\gamma}^{-25}\right\}_{k=-\infty}^{\infty}$ (d).}
\end{figure}
We compute for the eigenfunctions and eigenvalues of the GTO using the methods presented in reference \cite{caballar} to compute for the eigenvalues and eigenfunctions of the CTO. Specifically, we use the energy representation of the GTO, which is an infinite matrix whose nondiagonal elements are identical to the nondiagonal elements of the energy representation of the CTO. However, the diagonal elements of the energy representation of the GTO will have the explicit form
\begin{equation}
\left(\mathbf{T}_{\gamma,\alpha,2}\right)_{j,j}=\alpha_{k}
\end{equation}
which are different from the diagonal elements of the energy representation of the CTO, which are all equal to zero. We then truncate the resulting infinite matrix and compute for the eigenfunctions and eigenvalues of the resulting finite truncated matrix. We are assured that the resulting eigenfunctions and eigenvalues of this truncated matrix are numerically close to the actual eigenvalues $\tau_n$ and eigenfunctions $\ket{\varphi_{n,\gamma,\nu}^{0,2}}$ of the GTO, since the GTO is a compact operator \cite{taylor}.
\subsubsection{Time Evolution of the GTO Eigenfunctions}
Once we are able to compute for the eigenfunctions of the GTO, we then proceed to evolve these eigenfunctions over time, and we utilize the transition probability, whose explicit form is given by equation \ref{transprob}, in order to determine the probability that a time-evolved GTO eigenfunction $U_{t}\ket{\varphi_{n,\gamma,\nu}^{0,2}}=\ket{\varphi_{n,\gamma,\nu}^{0,2}(t)}$ will make a transition to another GTO eigenfunction $\ket{\varphi_{n,\gamma,\nu}^{0,2}}$. We then compare the resulting transition probabilities corresponding to the time-evolved GTO eigenfunctions with the transition probabilities corresponding to the time-evolved CTO eigenfunctions. The $\alpha$ sequence of the GTO to be considered in this section is $\alpha=\left\{E_{k,\gamma}^{-n}\right\}_{k=-\infty}^{\infty}$, with $n\geq 2$ and $E_{k,\gamma}$ the energy eigenvalues for the system, which is not $\tau$-symmetric. This $\alpha$ sequence was chosen in order to ensure that the kernel for the GTO will remain square integrable, thus preserving the self-adjointness and compactness of the GTO.

The results of our analysis of the dynamics of the GTO eigenfunctions via the transition probability are shown in figures 5 and 6. Figures 5b to 5d show that the plot of the transition probability generated by the time-evolved GTO eigenfunctions will most likely not have a single, well-defined peak which is localized at a single instant of time $t_{max}$, and if there is such a peak, the height of the peak will not be very close to or even equal to one. Furthermore, figures 6b to 6d show that the time when $P_{t}[n,n']$ will attain a maximum value, $t_{max}$, will not be close to, or even equal to, the difference between the eigenvalues $\tau_{n}-\tau_{n'}$ corresponding to the GTO eigenfunctions used to compute the transition probability, even if $\tau_{n}-\tau_{n'}\rightarrow 0$. As such, one cannot then physically interpret the GTO $\opr{T}_{2,\alpha,\gamma}$ whose corresponding $\alpha$ sequences will have nonzero elements in the same manner as the CTO $\opr{T}_{2,0,\gamma}$.
\section{Discussion}
We have shown in this work that there exist multiple solutions to the TE-CCR of the same category, which may be distinguished from each other via particular internal symmetries involving time reversal, in particular the symmetry which we call $\tau$-symmetry. We have also shown that the presence or absence of $\tau$-symmetry in particular elements of a given set of solutions to the TE-CCR of the same category can affect the dynamics of those solutions to the TE-CCR, thus affecting not just the manner by which we interpret those solutions to the TE-CCR via the system's internal unitary dynamics, but also ensuring that certain $\tau$-symmetric solutions to the TE-CCR of a given category will be physically distinct from their non $\tau$-symmetric counterparts within that category. 

This work, however, raises questions with regards to the precise role of the system's internal symmetries in formulating quantum time observables. In particular, how exactly are we to interpret non $\tau$-symmetric quantum time observables? More generally, is $\tau$-symmetry necessary in order to physically interpret quantum time observables? To be able to answer both of these questions, it is necessary to further investigate the dynamical behavior of the non $\tau$-symmetric ATO and GTO, in order for us to be able to determine whether or not one can still use the system's internal unitary dynamics to provide a physical interpretation to the ATO and GTO, and determine as well whether $\tau$-symmetry, or in general the internal symmetries of a system, is necessary in order to formulate solutions to the TE-CCR.

This work has been supported by a UP System Grant.

\end{document}